\documentclass{aa}

\usepackage{natbib}
\usepackage{url}

\usepackage{enumitem}

\usepackage{graphicx}
\usepackage{subfig}
\usepackage[varg]{txfonts}
\bibpunct{(}{)}{;}{a}{}{,} 

\usepackage{url}
\usepackage{hyperref}
\hypersetup{
     breaklinks,
     colorlinks   = true,
     citecolor    = blue,
     urlcolor     = blue,
     linkcolor    = blue
}

\begin{document}

   \title{Investigating the amplitude and rotation of the phase spiral \\in the Milky Way outer disc}

   \author{S. Alinder,
          P. J. McMillan 
          \and
          T. Bensby 
          }

   \institute{Lund Observatory, Division of Astrophysics, Department of Physics, 
            Lund University, Box 43, SE-221\,00 Lund, Sweden\\
              \email{simon.alinder@fysik.lu.se}
             }

    \date{Received March 31, 2023; Accepted July 9, 2023}
 
  \abstract
   { 
   With the data releases from the astrometric space mission \textit{Gaia}, the exploration of the structure of the Milky Way has developed in unprecedented detail and unveiled many previously unknown structures in the Galactic disc and halo. One such feature is the \textit{Gaia} phase spiral where the stars in the Galactic disc form a spiral density pattern in the $Z-V_Z$ plane. Many questions regarding the phase spiral remain, particularly how its amplitude and rotation change with position in the Galaxy. 
   }
   { 
   We aim to characterize the shape, rotation, amplitude, and metallicity of the phase spiral in the outer disc of the Milky Way. This will allow us to better understand which physical processes caused the phase spiral and can give further clues to the Milky Way's past and the events that contributed to its current state. 
   }
   { 
   We use \textit{Gaia} data release 3 (DR3) to get full position and velocity data on approximately 31.5 million stars, and metallicity for a subset of them. We then compute the angular momenta of the stars and develop a model to characterise the phase spiral in terms of amplitude and rotation at different locations in the disc.
   }
   { 
   We find that the rotation angle of the phase spiral changes with Galactic azimuth and Galactocentric radius, making the phase spiral appear to rotate about $3^\circ $ per degree in Galactic azimuth. Furthermore, we find that the phase spiral in the $2200 - 2400$\,kpc\,km\,s$^{-1}$ range of angular momentum is particularly strong compared to the phase spiral that can be observed in the solar neighbourhood. The metallicity of the phase spiral appears to match that of the Milky Way disc field stars.
   }
   { 
   We created a new model capable of fitting several key parameters of the \textit{Gaia} phase spiral. We have been able to determine the rotation rate of the phase spiral to be about $3^\circ $ per degree in Galactic azimuth. We find a peak in the amplitude of the phase spiral at $L_Z\,\approx\,2300$\,km\,kpc\,s$^{-1}$ which manifests as a very clear phase spiral when using only stars with similar angular momentum. 
   These results provide insights into the physical processes that led to the formation of the phase spiral and contribute to our understanding of the Milky Way's past and present state.
   }


   \maketitle
%

\section{Introduction}

How large spiral galaxies form and which processes contribute to their formation are open questions. By studying the structure of our own galaxy, the Milky Way, we can find traces of these processes and start to piece together its formation history. However, detailed structures that carry signatures of galaxy evolution and accretion events tend to phase mix and disappear with time. The outer disc of the Galaxy has longer dynamical timescales, meaning that dynamical and physical structures there remain for longer times \citep{freeman_new_2002}. Therefore, the outer Galactic disc is a good place to study when trying to answer questions about the Milky Way's past. 

The European Space Agency's \textit{Gaia} mission \citep{gaia_collaboration_gaia_2016} has provided accurate astrometric data for almost two billion stars in the Milky Way, and its different data releases (DR1 \citealt{gaia_collaboration_gaia_2016}, DR2 \citealt{gaia_collaboration_gaia_2018}, EDR3 \citealt{gaia_collaboration_gaia_2021}, and DR3 \citealt{gaia_collaboration_gaia_2023}) have allowed us to reveal ever more detailed and delicate structures in our Galaxy. Examples include the \textit{Gaia}-Enceladus-Sausage, the remnants of an ancient merger with a massive galaxy \citep{belokurov_co-formation_2018, helmi_merger_2018}; the Radcliffe wave, a large nearby structure of gas that contains several stellar nurseries \citep{alves_galactic-scale_2020}; the three-dimensional velocities of stars in the satellite dwarf galaxy Sculptor, allowing a close look at the kinematics of a dark matter dominated system \citep{massari_three-dimensional_2018}; many details about the structure of the Galactic halo leading to insights into its formation \citep{helmi_box_2017}; and the phase spiral (or ``snail shell''), a spiral pattern that was discovered by \cite{antoja_dynamically_2018} in the phase plane defined by the vertical distance from the Galactic plane ($Z$) and the vertical velocity component ($V_Z$). 
 
The existence of the phase spiral physically means that the distribution of the $V_Z$-velocities for the stars at certain $Z$-positions is uneven in a way that looks like a spiral when plotted on a phase space diagram. For example, when looking at stars in the solar neighbourhood with $Z \approx 0$\,pc, there are more stars with $V_Z \approx -20$\,km\,s$^{-1}$ and fewer stars with $V_Z \approx -15$\,km\,s$^{-1}$ than expected from a smooth symmetrical distribution. 
The phase spiral was mapped within a Galactocentric range of $7.2 < R/\,\mathrm{kpc} < 9.2$ and within $15^\circ$ of the anti-centre direction (opposite to the Galactic centre) by \cite{bland-hawthorn_galah_2019}, to $6.6 < R/\,\mathrm{kpc} < 10$ by \cite{laporte_footprints_2019}, to $6.34 < R/\,\mathrm{kpc} < 12.34$ by \cite{wang_galactic_2019}, and \cite{xu_exploring_2020} extended the furthest outer detection to 15\,kpc from the Galactic centre. When investigations and simulations of the phase spiral were done across a larger range of positions in the Galaxy, these studies found that the phase spiral changes shape with Galactocentric radius. Close to the solar radius, it has a greater extent in the $V_Z$ direction, and at greater Galactocentric radii it has a larger extent in the $Z$ direction. 
This increase in vertical extent at greater Galactocentric distances is due to the change in gravitational potential and a reduction in vertical restoring force. 

The phase spiral is thought to be a response of the Galactic  disc to a perturbation that pushed it out of equilibrium. This response, over time, winds up in the $Z$-$V_Z$~plane into a spiral due to phase-mixing. In this simple picture, the time since the perturbation determines how wound the phase spiral has become, while any variation with Galactic azimuth, such as a rotation of the phase spiral in the $Z$-$V_Z$~plane, corresponds to a difference in the initial perturbation felt by stars at different azimuths.  
\cite{wang_galactic_2019} looked at the phase spiral at different Galactic azimuths and found that the amplitude of the spiral pattern changes. 
\cite{widmark_weighing_2022} show that the orientation of the phase spiral changes with Galactic azimuth and that the difference across $180^\circ$ of the Galactic azimuth in a heliocentric system will be about $140^\circ$. They show a very slight positive change in angle with radial distance, but only in cells they have marked as less reliable (see \citealt{widmark_mapping_2022}, Figs. D.1 and D.2 for details). 
\cite{bland-hawthorn_galactic_2021} show the rotation of the phase spiral at different Galactic azimuths in their N-body simulation of the effects of the passage of the Sagittarius dwarf galaxy on the Galactic disc.
\cite{darragh-ford_$textttescargot$_2023} show the rotation of the phase spiral at different Galactic azimuths and angular momenta in their model.
The rotation of the phase spiral is an important part of any attempt at modelling it directly, and an important property to capture in any simulation because it is tied to the potential of the Galactic disc. In this study, we will present measurements of the propagation of the rotation angle of the phase spiral. 

The chemical composition of the phase spiral was investigated by \cite{bland-hawthorn_galah_2019} using elemental abundances from the GALAH survey \citep{buder_galah_2018}. They found no evidence that the phase spiral is a single-age population (such as a star cluster or similar) because the trend in metallicity is smoothly varying. This indicates that the stars in the phase spiral are part of the general population of the Milky Way disc. \cite{an_asymmetric_2019}, using data from APOGEE DR14 \citep{abolfathi_fourteenth_2018}, examined the metallicity of the Galactic disc and found an asymmetry in the $Z$-direction, with higher mean metallicity above the plane of the Galaxy than below. They explain this asymmetry as being caused by the phase spiral as it would push stars to greater $ Z$-distances. These results are reported as being in agreement with the findings of \cite{bland-hawthorn_galah_2019}. 
In this study, we use global metallicity data on a large number of stars to investigate the chemical properties of the phase spiral. 

Several theories for the origin of the phase spiral exist in the literature. Among the proposed scenarios, the most popular one is that the phase spiral was caused by gravitational interactions between the Milky Way and a massive external object. The primary observational evidence for this scenario is the presence of the Sagittarius dwarf galaxy \citep{ibata_dwarf_1994} which is undergoing disruption by the Milky Way \citep{binney_origin_2018, laporte_footprints_2019, bland-hawthorn_galah_2019}. If the Sagittarius dwarf galaxy is the cause, then the properties of the phase spiral and the properties of the Sagittarius dwarf galaxy at the time when the interaction took place are linked and knowledge of one can be used to derive the properties of the other, for example, the mass history of the Sagittarius dwarf galaxy, and the time of impact \citep{bland-hawthorn_galactic_2021}. 
\cite{darling_emergence_2019} discusses the possibility that the phase spiral is caused by bending waves (physical displacement of stars). Several phenomena can cause these waves, including dwarf galaxy impacts and gravitational effects from the bar or spiral structure of the Galaxy. 
\cite{frankel_vertical_2023} and \cite{antoja_phase_2023} both find that a simple model with a single cause for the perturbation fails to explain the observations and calls for more complex models. 
\cite{hunt_multiple_2022, bennett_exploring_2022} and \cite{tremaine_origin_2023} suggest, in different ways, that the formation history of the phase spiral cannot be explained with a single impact but perhaps rather originates from several small disturbances. 

The primary goal of this paper is to map the rotational angle, amplitude, and chemical composition of the phase spiral. By using the most recent data from \textit{Gaia}, DR3, we aim to investigate these properties in higher definition than before. 
As we learn more about the extent, amplitude, rotation, and shape of the phase spiral, we might be able to strengthen the evidence for one of the proposed formation scenarios, leading to a greater understanding of the formation history of the Milky Way. 
The model we create is independent of any particular narrative or formation scenario. It is based solely on observations. This is a deliberate strategy aimed at reducing the risk that assumptions about what the phase spiral is will affect our conclusions. 
As shown by \cite{widrow_swing_2023}, authors should be careful when trusting simple kinematic models because the self-gravity of the perturbation is significant for its evolution. We can see this realized in papers which report mutually incompatible results. For example, the recent papers by \cite{frankel_vertical_2023} and \cite{darragh-ford_$textttescargot$_2023}, who disagree on the time since the impact that formed the phase spiral with the former claiming that ``there is no well-defined global dynamical age of a single perturbation'' and the latter presents results that ``range from 288 to 966 Myrs with a median of 524 Myrs''.
We start by presenting how the stellar sample is selected in Sect.~\ref{section:data}. 
In Sect.~\ref{section:spiral} we develop the model that we use to analyse the phase spiral and how it changes across the Galactic disc. 
In Sect.~\ref{ssection:chem} we examine the chemical composition of the phase spiral and in Sect.~\ref{section:discussion} we discuss our results. Finally, we summarise our findings and conclusions in Sect.~\ref{section:conclusions}. 

\section{Data}\label{section:data}
\begin{figure*}[ht]
\includegraphics[width=\hsize]{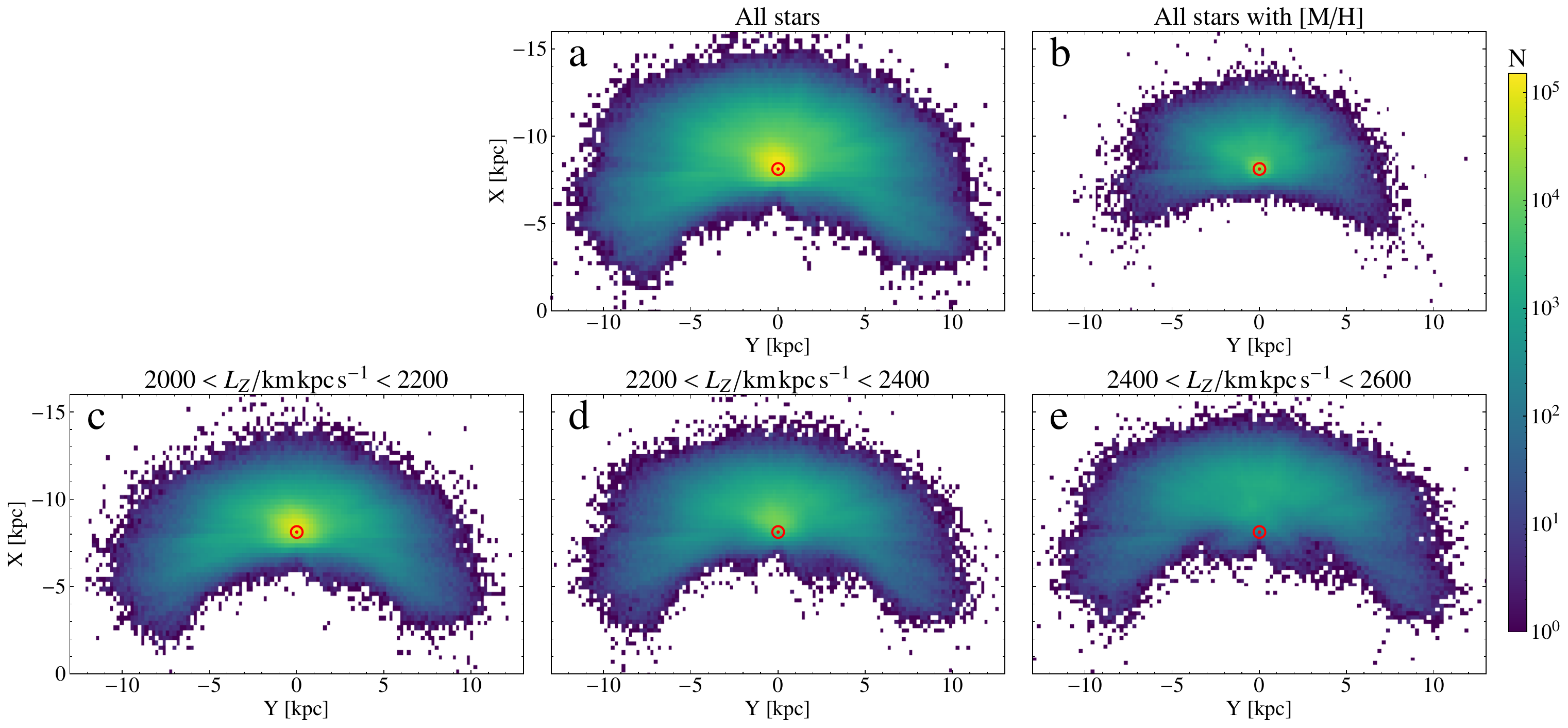}
  \caption{Spacial distribution of stars in the data. Top middle panel: Number density of stars used in our investigation in the Galactocentric Cartesian $X$-$Y$ plane. This panel contains stars in the $2000 < L_Z/$\,kpc\,km\,s$^{-1} < 2600$ range. Top right panel: Number density of stars within our sample that have global metallicity [M/H] values.
  Bottom panels: Number density of the selected stars in the used angular momentum bins in the $X$-$Y$ plane. The circled red dot is the location of the Sun in all panels. 
  The bin size for all panels is 200\,pc by 200\,pc.
          } 
     \label{fig:data_location}
\end{figure*}

To study the phase spiral we need stars with known three-dimensional velocities. We use \textit{Gaia} DR3 \citep{gaia_collaboration_gaia_2016, gaia_collaboration_gaia_2023} to get positions, proper motions, and radial velocities for the stars. The distances were calculated by \cite{bailer-jones_estimating_2021} who used a Bayesian approach with a direction-dependant prior on distance, the measured parallaxes, and \textit{Gaia} photometry, exploiting the fact that stars of different colours have different ranges of probable absolute magnitudes. The ADQL-query used to retrieve this data from the public \textit{Gaia} database\footnote{\url{https://gea.esac.esa.int/archive/}} was:
\begin{verbatim}
SELECT source_id, ra, dec, pmra, pmdec, 
r_med_photogeo, radial_velocity 
FROM external.gaiaedr3_distance
JOIN gaiadr3.gaia_source USING (source_id)
WHERE parallax_over_error>=3 
and radial_velocity IS NOT NULL 
and r_med_photogeo IS NOT NULL
\end{verbatim}
This query resulted in 31\,552\,449 stars being selected. We use \verb|parallax_over_error>=3 | as this removes the most uncertain distance measurements. 

For the chemical investigation, we use the global metallicity [M/H] data from \textit{Gaia} DR3 RVS spectra \citep{recio-blanco_gaia_2023} with the ADQL-query: 
\begin{verbatim}
SELECT source_id, mh_gspspec, flags_gspspec 
FROM gaiadr3.astrophysical_parameters
JOIN gaiadr3.gaia_source USING (source_id)
WHERE parallax_over_error>=3 
and teff_gspspec > 3500
and logg_gspspec BETWEEN 0 and 5
and teff_gspspec_upper - teff_gspspec_lower < 750
and logg_gspspec_upper - logg_gspspec_lower < 1
and mh_gspspec_upper - mh_gspspec_lower < 5
and mh_gspspec IS NOT NULL 
and radial_velocity IS NOT NULL 
\end{verbatim}
This query resulted in 4\,516\,021 stars being selected. We use quality cuts as recommended by \cite{recio-blanco_gaia_2023} combined with those used for the main sample. These cuts remove stars with low temperatures as these stars are known to have complex, crowded spectra and stars with $\log(g)$ and $T_{\mathrm{eff}}$ not between the upper and lower confidence levels. We also filter out the least reliable K and M-type giant stars using the supplied flag as there exists a parameterisation problem for cooler and metal-rich stars of this type. The final sample for the chemical investigation consists of this table combined with the previous one to get positions, velocities and spectral data in the same sample and is 4\,303\,484 stars after quality cuts. 

We use a Galactocentric coordinate system centred on the Galactic centre with the Sun on the (negative) $X$-axis at a distance of 8.122\,kpc and a height of 20.8\,pc with the $Y$-axis pointing towards $l=90^{\circ}$ and the $Z$-axis pointing to $b=90^{\circ}$. Galactic azimuth ($\phi$) is decreasing in the direction of Galactic rotation and the Sun is located at $\phi = 180^\circ$. The velocity of the Sun is $[V_{R, \sun}=-12.9, V_{\phi, \sun}=-245.6, V_{Z, \sun}=-7.78]$ km\,s$^{-1}$ \citep{reid_proper_2004, drimmel_solar_2018, gravity_collaboration_detection_2018, bennett_vertical_2019}. For the computations and definitions of coordinates, we use  Astropy v5.2 \citep{astropy_collaboration_astropy_2022}.
For reasons given in Sect.~\ref{section:spiral}, we will base our analysis on samples defined by the angular momenta of the stars. The angular momentum is computed as $ L_Z = R \,|V_{\phi}|$.

The distribution of the stars in the Galactocentric Cartesian $X$-$Y$ plane is shown in Fig.~\ref{fig:data_location}. 
We can see that the sample mostly contains stars with Galactocentric distances $5-12$\,kpc. This allows us to study the phase spiral in regions far from the solar neighbourhood and measure how it changes with location. The top row shows the full sample to the left and the sample of stars with [M/H] values to the right. The bottom row is split into three bins with different angular momentum. In the bin with the highest angular momentum (right-most panel), most stars are ${\sim} 2$\,kpc further out than those in the low  angular momentum bin (left-most panel). 

\begin{figure*}
\centering
\includegraphics[width=0.8\hsize]{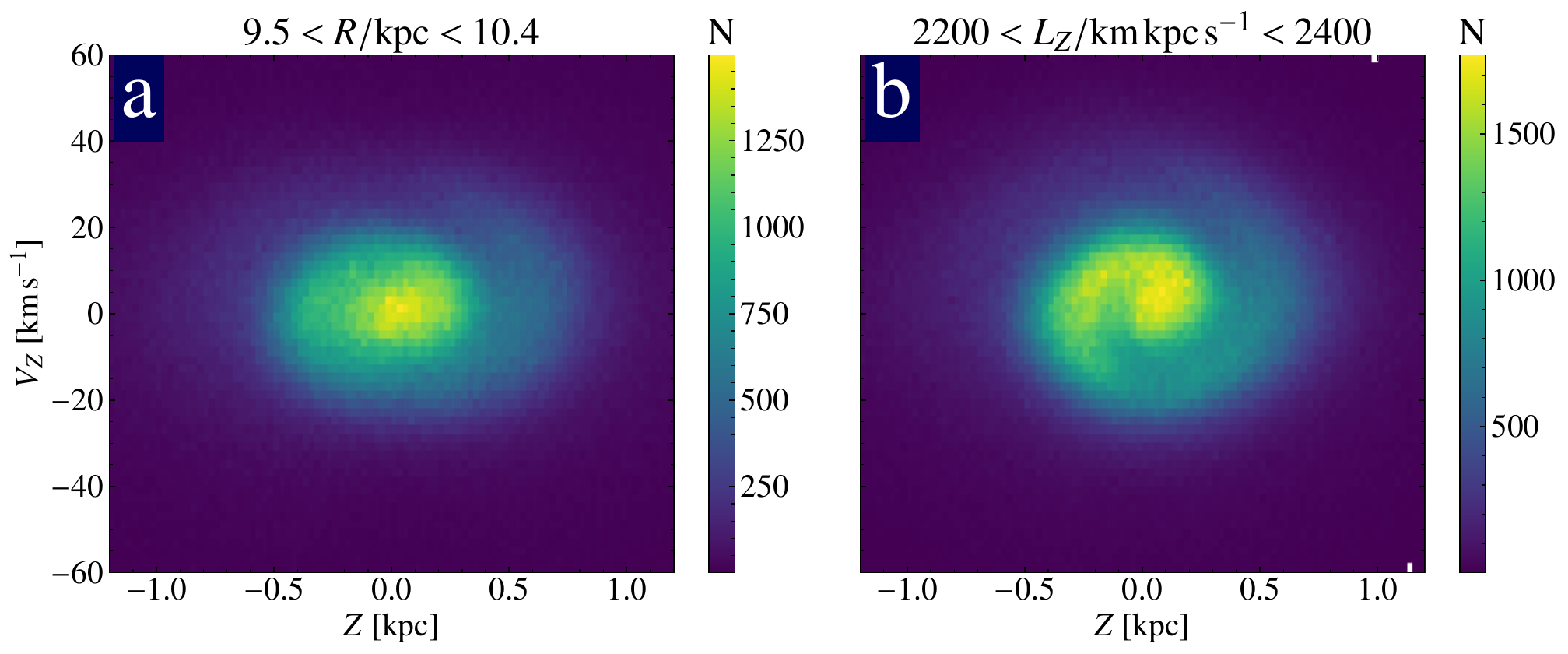}
  \caption{Comparison of phase spirals made of stars selected using different methods. {\sl a:} The number density of stars in the $Z$-$V_Z$ phase plane in the $9.5 < R /$ kpc $ < 10.4$ range. {\sl b:} The number density of stars in the $Z$-$V_Z$ phase plane in the $2200 < L_Z/$\,kpc\,km\,s$^{-1} < 2400$ range.
  Panel {\sl a} shows a less clearly defined phase spiral than the one in panel {\sl b}. 
          } 
     \label{fig:new_v_old_style}
\end{figure*}

For an investigation of a structure of stars like the phase spiral, a velocity-dependant selection will produce a more sharply defined phase spiral than a position-dependent selection because the phase spiral is a dynamical structure \citep{bland-hawthorn_galah_2019, li_vertical_2021, antoja_phase_2023}. As noted by \cite{hunt_resolving_2021} and \cite{gandhi_snails_2022}, samples based on position will contain stars with a wide range of velocities and orbital frequencies because stars with different guiding centre radii will temporarily be close together. This means one will indirectly sample a large part of the Galaxy, which can be useful when addressing other research questions. An example of this is in \cite{bensby_exploring_2014} where relatively nearby stars were sampled to map the age and abundance structure of several components of the Milky Way. However, for our purposes, it is more meaningful to group stars in terms of their dynamic properties rather than position. 

Here we do a quick comparison of samples selected by Galactocentric position and by angular momentum. Using the Galactic potential from \cite{mcmillan_mass_2017}, we compute the guiding centres for hypothetical stars with $L_Z = 2200$\,kpc\,km\,s$^{-1}$ and $L_Z = 2400$\,kpc\,km\,s$^{-1}$ to be $R_g \approx 9.5$\,kpc and $R_g \approx 10.4$\,kpc, respectively. The $Z $-$ V_Z$ phase space for the stars between the Galactocentric radii corresponding to these guiding radii are shown in Fig.~\ref{fig:new_v_old_style}a, and the same for stars in the angular momentum range in Fig.~\ref{fig:new_v_old_style}b. The phase spiral based on stars in the $9.5 < R\,/$ kpc $< 10.4$ range is visible but less clear while the phase spiral based on stars in the $2250 < L_Z\,/$ kpc\,km\,s$^{-1} < 2350$ range is more prominent. This is because Fig.~\ref{fig:new_v_old_style}a contains stars that are part of different-looking phase spirals, making the stars fill in the gaps in each other's patterns, whereas Fig.~\ref{fig:new_v_old_style}b mostly contains stars that are part of one single phase spiral, making the pattern clear. Figure~\ref{fig:new_v_old_style}a contains a total of 1\,045\,921 stars, while Fig.~\ref{fig:new_v_old_style}b contains 1\,348\,886 stars.

\section{The Gaia phase spiral}\label{section:spiral}
\begin{figure*}
\centering
\includegraphics[width=0.9\hsize]{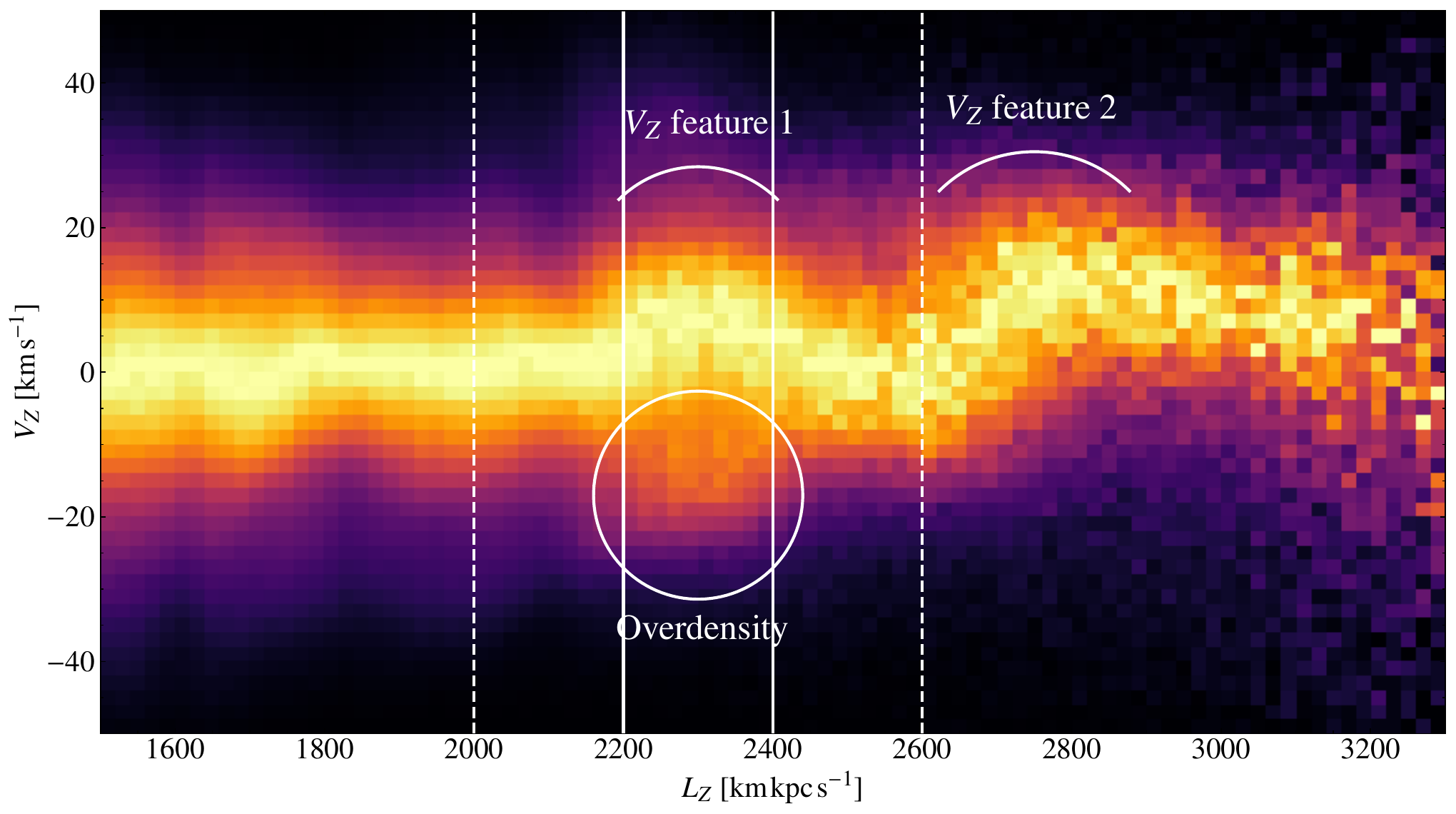}
   \caption{Column normalized histogram of star number density in the $L_Z-V_Z$ plane in the Galactic outer disc. 
   The region of interest is marked by solid white lines at $L_Z = [2200, 2400]$\,kpc\,km\,s$^{-1}$, with dashed white lines at $L_Z = 2000 $ and $2600$\,kpc\,km\,s$^{-1}$ marking the areas used for comparisons in Figs. \ref{fig:data_location}, \ref{fig:phase_spiral_and_alpha_LZ}, and \ref{fig:metal_spiral}. Features mentioned in the text are also marked. 
   The figure contains all stars in our sample with $175^\circ < \phi < 185^\circ$, 12\,723\,513 in total. 
   } 
   \label{fig:L-plot}
\end{figure*}

Figure~\ref{fig:L-plot} shows the density of stars within $5^\circ$ of the anti-centre direction plotted in the $L_Z$-$V_Z$ plane. The thick line is the Galactic disc and the ``$V_Z$ feature 2'' at $L_Z\,\approx\,2700$\,kpc\,km\,s$^{-1}$ is the bifurcation which was discovered by \citet{gaia_collaboration_gaia_2021} and investigated by \cite{mcmillan_disturbed_2022} who found that it may be an effect of the passage of the Sagittarius dwarf galaxy. Several other features can be seen, but a particularly clear one that we focus on is the wrinkle labelled ``$V_Z$ feature 1'' at $L_Z\approx 2300$\,kpc\,km\,s$^{-1}$ and the apparent overdensity centred on $(L_Z, V_Z) = (2300\,\mathrm{kpc}\,\mathrm{km}\,\mathrm{s}^{-1}, -20\,\mathrm{km}\,\mathrm{s}^{-1})$. 
These regions and features are marked lines in Fig.~\ref{fig:L-plot}. Finding this seemingly isolated overdensity of stars sitting below the thick line was surprising since the stars otherwise show a smooth falloff from the centre in the vertical directions. 
As we will show, the cause for the highlighted overdensity and $V_Z$ feature 1 in Fig.~\ref{fig:L-plot} seems to be that, in the range $2200 < L_Z/$\,kpc\,km\,s$^{-1} < 2400$, a higher proportion of the stars are part of the phase spiral, giving the stars in that region an unusual $V_Z$ distribution and a very prominent phase spiral. 

\subsection{Model of the phase spiral} \label{ssection:model}

\begin{table*}
\centering
\caption{Free parameters in the model of the phase spiral.}
\begin{tabular}{ccccc}
\hline \hline
Name & min & max & Description & Unit \\
\hline
$\alpha$ & 0 & 1 & Amplitude of spiral pattern & -- \\
$b$ & 0.01 & 0.175 & Linear winding parameter & $\mathrm{pc\,rad^{-1}}$ \\
$c$ & 0.0 & 0.005 & Quadratic winding parameter & $\mathrm{pc\,rad^{-2}}$ \\
$\theta_0$ & $-\pi$ & $\pi$ & Angle offset & rad \\
$S$ & 30 & 70 & Scale factor & $\mathrm{km\,s^{-1}\,kpc^{-1}}$ \\
$\rho$ & 0 & 0.3 & Flattening function distance & kpc \\
\hline
\end{tabular}
\label{tab:priors} 
\tablefoot{For each parameter we assume a prior probability that is constant between these min and max values.}
\end{table*}

\begin{figure}
\centering
\includegraphics[width=\hsize]{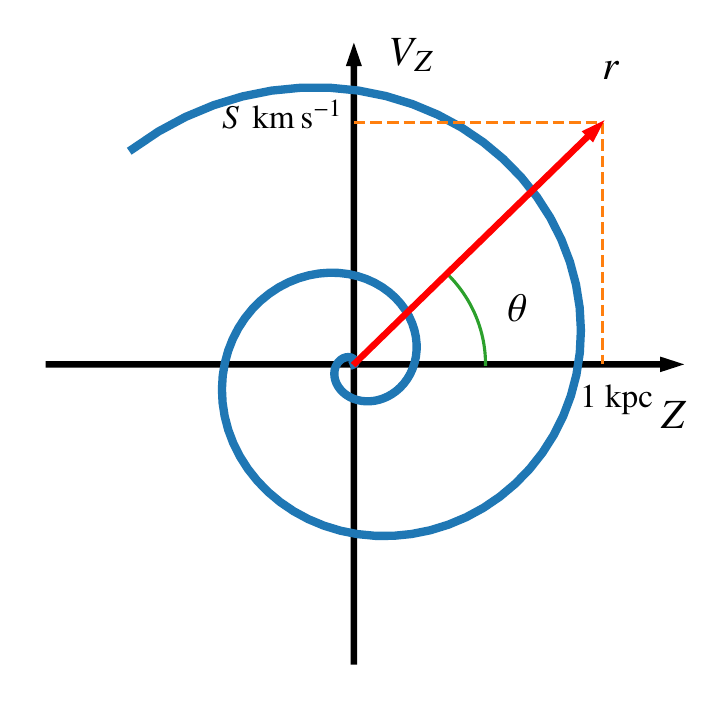}
  \caption{Illustration of the phase-plane coordinates used. $r$ is the phase distance and $\theta$ is the phase angle. In this example, $\theta = 45^\circ$. The scale factor $S$ has been chosen such that the $r$-vector could be drawn with constant length regardless of angle. 
          }
     \label{fig:spiral_coords}
\end{figure}

\begin{figure}
\centering
\includegraphics[width=\hsize]{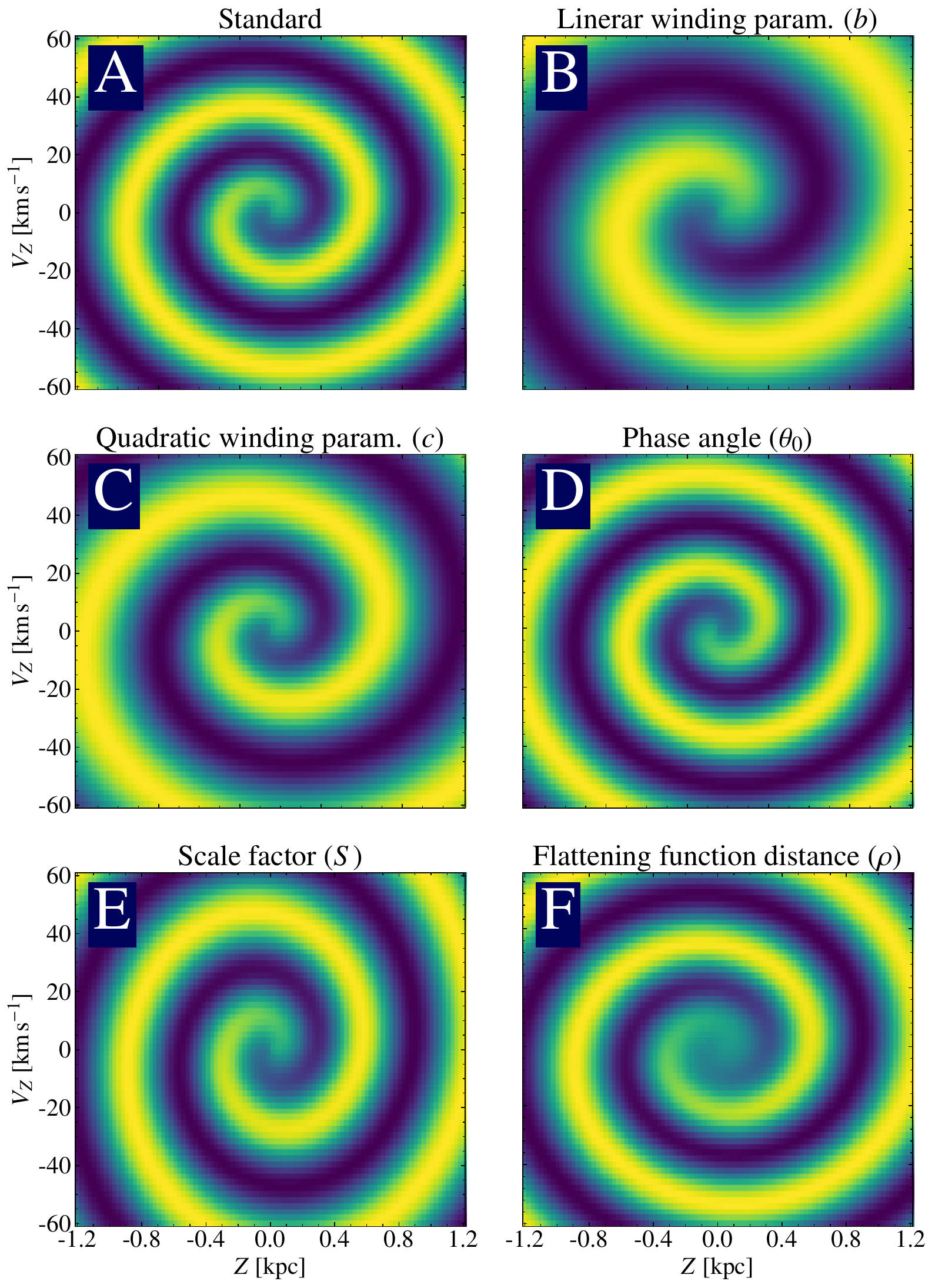}
  \caption{Examples of the effects on the spiral perturbation when changing (increasing) the different parameters in the model.
  Panel A shows the spiral perturbation for a certain set of parameters. This is taken as the default for the comparison in this figure. Panel B shows the spiral perturbation with an increased linear winding parameter. Panel C shows the spiral perturbation with an increased quadratic winding parameter. Note that the inner part of the spiral is still similar to panel A. Panel D shows the spiral perturbation with an increased phase angle, rotating it half a revolution. Panel E shows the spiral perturbation with an increased scale factor which increases the $V_Z$-$Z$ axis ratio. Panel F shows the spiral perturbation with the flattening function distance increased which makes the inner parts less distinct. 
          }
     \label{fig:params_demo}
\end{figure}

To quantify the properties of the phase spiral as functions of $R, \phi$ and $L_Z$ we construct a model inspired by those used by \cite{widmark_weighing_2021} and \cite{guo_northsouth_2022}. The model is built by creating a smoothed background from observed data, then a spiral-shaped perturbation that, when multiplied by the background, recreates the observed distribution of stars in the $Z$-$V_Z$ plane. This way, the spiral can be isolated and quantified. In this model, we compute values for the phase distance $r$ and the phase angle $\theta$ by using 
\begin{equation}
\begin{split}
    r = r(Z, V_Z) = \sqrt{Z^2 + \left(\frac{V_Z}{S}\right)^2},
\end{split}
\label{eqn:r}
\end{equation}

\begin{equation}
\begin{split}
    \theta = \theta(Z, V_Z) = \arctan\left(\frac{1}{S}\frac{V_Z}{Z} \right),
\end{split}
\label{eqn:theta}
\end{equation}
where $S$ is a scale factor which determines the ratio between the axes and is a free parameter in the model. These coordinates are illustrated in Fig.~\ref{fig:spiral_coords} with a simple diagram. A larger value of $S$ stretches the $V_Z$ axis, thus controlling the axis ratio of the spiral, see panel E in Fig.~\ref{fig:params_demo}. 
A star experiencing harmonic motion in $Z$ will trace a circle in the phase plane for the right value of $S$, since $S$ is closely related to the vertical frequency of oscillations in the Galactic disc. We restrict it to $30 < S < 70$, where $S$ is in units of $\mathrm{km\,s^{-1}\,kpc^{-1}}$, as this is the range stars tend to be in.

Starting from \cite{guo_northsouth_2022}'s discussion of the shape of the phase spiral, we consider the quadratic spiral
\begin{equation}
    r = a + b \phi_s + c \phi_s^2,
\end{equation}
where $\phi_s$ is the angle of the spiral. They claim that an Archimedean spiral\footnote{An Archimedean spiral is a spiral with a linear relation between angle and radial distance. Expressed in polar coordinates the spiral traces $r = b\phi_s$.} ($a = 0, c = 0$) fits the data well enough. We, however, found that our model fits better when we do not require that $c = 0$. We can assume $a = 0$ without loss of generality. As we construct the model we will be referring to Fig.~\ref{fig:params_demo} for illustrations of the effects of each parameter. Figure~\ref{fig:params_demo} contains six panels. Panel A shows the spiral perturbation for a certain set of parameters. Each other panel shows the spiral perturbation with one parameter increased and will be referred to as that parameter is introduced. 
We write the equation for the radial distance of the phase spiral as
\begin{equation}
    r = b \phi_s + c \phi_s^2,
\label{eqn:}
\end{equation}
which means
\begin{equation}
    \phi_s(r) = -\frac{b}{2 c} + \sqrt{\left(\frac{b}{2 c}\right)^2+\frac{r}{c}}.
\label{eqn:quadratic}
\end{equation}
The parameter $b$ is the linear winding term of the spiral. Higher values of $b$ mean the spiral winds slower and moves further in $r$ per turn, see the top-middle panel in Fig.~\ref{fig:params_demo}. The value of $b$ has to be positive and by inspection, we find it to provide reasonable results between 0.01 and 0.175.
$c$ is the quadratic winding term. It has a similar meaning to $b$ except it does not act equally at all $r$, having a smaller effect close to the middle of the spiral and a greater further out, see the top-right panel of Fig.~\ref{fig:params_demo}. $c = 0$ means that the spiral is Archimedean and its radius has a constant increase with increasing angle. $c$ has to be positive and, by inspection, we find that by limiting its upper value to $0.005$ we get reasonable results. 

Following \cite{widmark_weighing_2021} we take the form of the perturbation to be 
\begin{equation}
    f(r, \Delta \theta) = 1 + \alpha \cdot g(r) \cos(\Delta \theta),
\label{eqn:basis_of_model}
\end{equation}
where $\alpha$ is a free parameter of the model that defines the amplitude of the phase spiral. This spiral perturbation can have values in the range $1+\alpha $ to $ 1-\alpha$. If $\alpha = 0$ then the smoothed background is unperturbed by the spiral, if $\alpha = 1$ there are no stars that are not part of the spiral. We define $\Delta \theta$ as the phase angle relative to the peak of the perturbation as a function of phase distance as
\begin{equation}
    \Delta \theta = \theta - \phi_s(r) - \theta_0,
\label{eqn:cos_angle}
\end{equation}
where $\theta_0$ is the angle offset, which is a free parameter, giving us
\begin{equation}
    f(r, \theta) = 1 + \alpha \cdot g(r) \cos(\theta - \phi_s(r) - \theta_0).
\label{eqn:form_of_model}
\end{equation}

\begin{figure}
\centering
\includegraphics[width=\hsize]{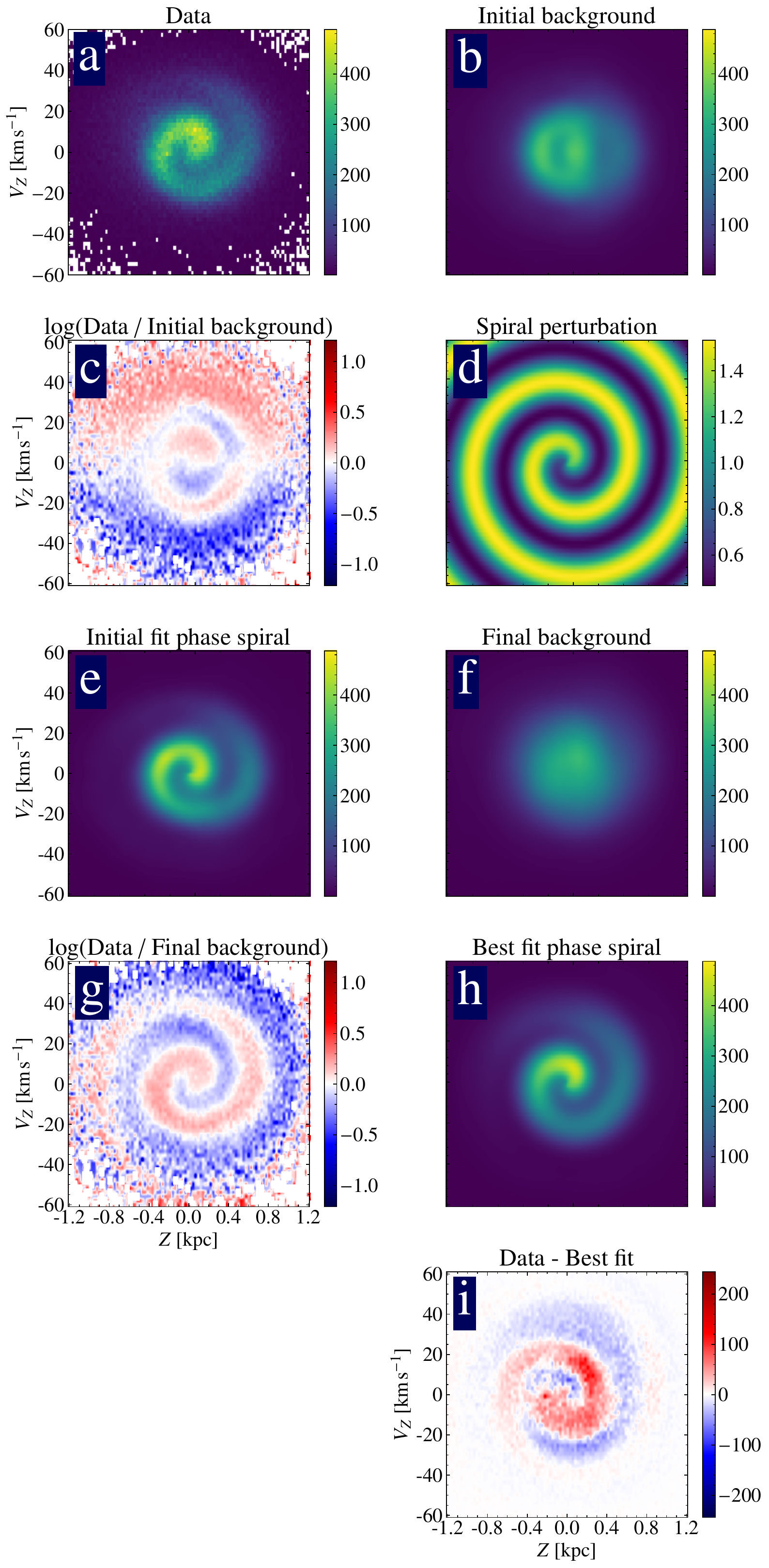}
  \caption{Example of the process from data to fitted model. 
  a: Data used for the model, a two-dimensional histogram showing number density consisting of 1,396,320 stars, b: Initial background, c: Data / initial background, d: Extracted spiral perturbation, e: Initial fit spiral. f: Final background, g: Data / final background, h: Best fit spiral. i: Residuals as computed as Data (panel a) - Best fit (panel h). See text for details on individual panels. 
  This example consists of stars with $8.4 < R / \mathrm{kpc} < 10.4$ and $ 165^\circ < \phi < 195^\circ$. 
          }
     \label{fig:spiral_model_demo}
\end{figure}

\begin{figure}
\centering
\includegraphics[width=\hsize]{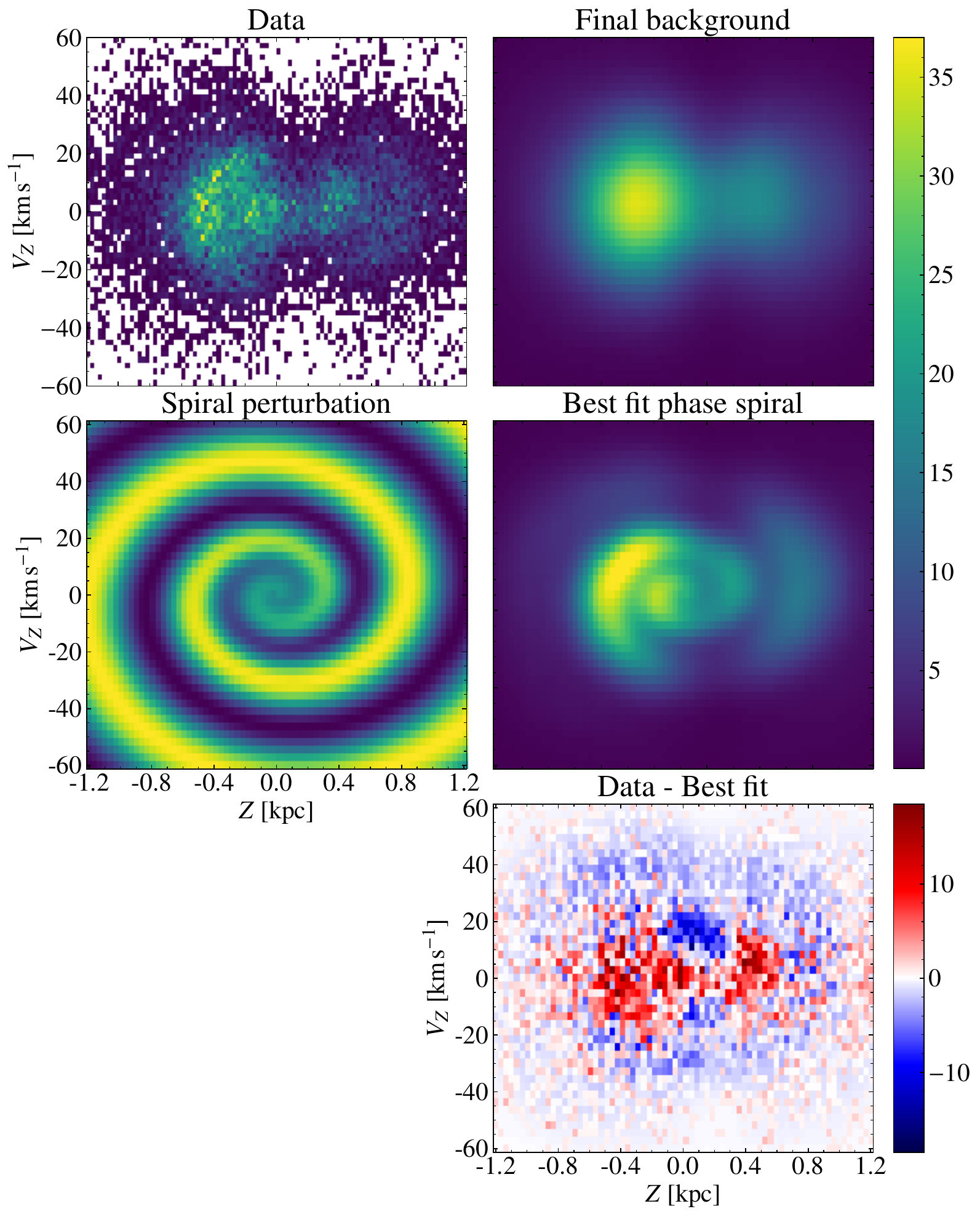}
  \caption{Example of a selection of stars near the edge of our considered area, containing only about 22,000 stars. 
  The sample contains stars with $8 < R / \mathrm{kpc} < 12$ and $ 150^\circ < \phi < 155^\circ$.
  Upper left: The phase plane showing strong extinction by dust.
  Upper right: The background produced by the model.
  Lower left: The spiral perturbation produced by the model (this panel does not share the colour bar with the rest).
  Lower right: The best fit. 
  Bottom: Residuals computed as Data - Best fit.
  We can see that even absent a clear spiral pattern in the data, the model still produces a convincing spiral and fit. 
          }
     \label{fig:bad_example}
\end{figure}

\begin{figure*}
\centering
\includegraphics[width=\hsize]{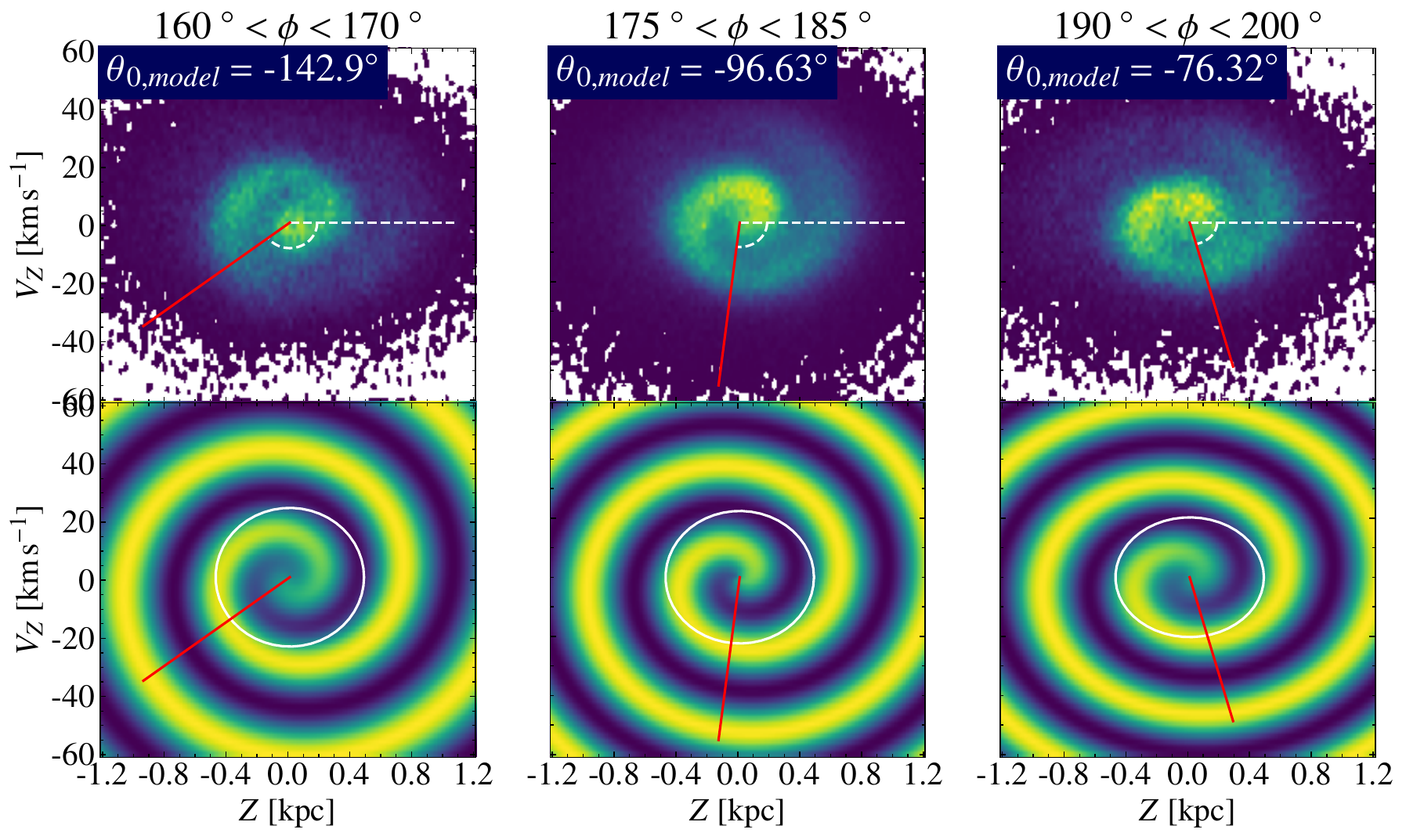}
  \caption{Demonstration of the rotation of the phase spiral with visualisation of how the model found the angles. Upper row: The phase spiral at low, medium, and high Galactic azimuth ($\phi$) with the angle $\theta_{\,0,\:\mathrm{model}}$ marked with a red line and $\theta_{\,0,\:\mathrm{model}} = 0$ marked with a white dashed line. Lower row: The corresponding spiral perturbations fitted to the data with $\theta_{\,0,\:\mathrm{model}}$ marked with a red line and the measurement distance for $\theta_{\,0,\:\mathrm{model}}$ marked with a white ring. 
          }
     \label{fig:spiral_spin}
\end{figure*}

The $g(r)$ term in Eqs. \ref{eqn:basis_of_model} and \ref{eqn:form_of_model} represents a flattening function that we will define here.
The innermost part of the phase spiral cannot be accurately fitted with this kind of model since the part of the Galactic disc with low $ Z$-displacement and velocity is subject to small perturbations which wash out the phase spiral. We therefore apply a flattening function called $g$ to reduce the strength of the spiral perturbation in this region. Like \cite{widmark_weighing_2021}, we use the logistic function (a sigmoid function) for our flattening function. The logistic function has the property that it is bounded by zero and one, and smoothly (exponentially) changes between them, thereby bringing any value into the zero-to-one range in a naturalistic way. We define the flattening function as
\begin{equation}
    g(r) = \mathrm{sigm}\left(\frac{r - \rho}{0.1\,\mathrm{kpc}}\right),
\label{eqn:inner_mask}
\end{equation}
where 
\begin{equation}
    \mathrm{sigm}(x) = \frac{1}{1+e^{-x}},
\label{eqn:sigmoid}
\end{equation}
is the sigmoid function and $\rho$ is the radius parameter of the function, which is a free parameter in the model. The flattening function reduces the impact of the inner part of the spiral by ``flattening'' it, bringing it closer to one, see the bottom-right panel of Fig.~\ref{fig:params_demo}. A larger value of $\rho$ means a larger part of the spiral is flattened. By inspection, we find that this value should be less than 0.3 which we apply as a prior. 

We also want to reduce the statistical influence of the most distant regions of the phase plane where there are very few stars. Similarly to \cite{widmark_weighing_2021}, we multiply both data and model by a term that we call mask. We define it as
\begin{equation}
    \mathrm{mask}(Z, V_Z) = -\mathrm{sigm}\left(\left( \frac{Z}{1 \,\mathrm{kpc}} \right)^2 + \left( \frac{V_Z}{40 \,\mathrm{km\,s}^{-1}}\right)^2 - 1 \right) -1.
\label{eqn:outer_mask}
\end{equation}
This mask is applied when evaluating how good the fit is. Mask is the only term in our model which originates in data analysis, not some physical property of the phase spiral. We do this because we want the model to correspond to real features. Removing the stars in the outer regions of the phase plane is somewhat arbitrary but justified because those regions are very sparse and therefore will only contribute noise. 

Combining Eqs.~\ref{eqn:quadratic}, \ref{eqn:form_of_model}, and \ref{eqn:inner_mask} gives the spiral perturbation as
\begin{equation}
    f(r, \theta) = 1 + \alpha \cdot \mathrm{sigm}\left(\frac{r - \rho}{0.1\,\mathrm{kpc}}\right) \cos(\theta - \phi_s(r) - \theta_0),
\label{eqn:spiral}
\end{equation}
where $\alpha, \rho$, and $\theta_0$ as well as $b$ and $c$ are free parameters of the model. 
The prior we use is based on observations of the phase spiral and chosen in a way to ensure that the sampler converges to the most reasonable solution. The prior uses uniform probabilities for all parameters between the values listed in Table~\ref{tab:priors}. This table also contains a summary of the parameters with their units. 

\subsection{Fitting procedure}
The spiral perturbation is found through an algorithm that involves an iterative procedure to create a smooth background ($B$). With a smooth background, we can define the perturbation which has the parameters of the phase spiral. This background is complicated and changes depending on where in the Galaxy you look, in part because interstellar extinction hides stars in the plane of the Galactic disc at greater distances creating a vertical region of lower number density in the middle of the phase plane (see Fig.~\ref{fig:bad_example}). 
Our previous methods attempted to fit a two-dimensional Gaussian to the data for use as a background but this performed poorly in regions with low star counts. A precursor to the method ultimately used involved creating a background based on the data as described below but without any iterative refinement. This, however, led to a degree of numerical instability in the results that was resolved by using this iterative method instead. 

The procedure for fitting a spiral perturbation to the data is illustrated in Fig.~\ref{fig:spiral_model_demo} and the letters in this subsection refer to the panels in the figure. The figure contains stars with $8.4 < R /$ kpc $ < 10.4$ and, $165^\circ < \phi < 195^\circ$ in the $2200 < L_Z/$\,kpc\,km\,s$^{-1} < 2400$ range. The procedure contains the following steps. 
\begin{enumerate}
    \item Collect the data into a two-dimensional number density histogram in the $Z-V_Z$ phase plane (panel a). The model uses a bin size of 25\,pc by 2 km\,s$^{-1}$ except in cases with fewer stars when larger bins are used. For example, Fig.~\ref{fig:bad_example} uses bins of $33 \frac{1}{3}$\,pc by $2 \frac{2}{3}$\,km\,s$^{-1}$. 
    \item Create the first background using the observed data (panel b). The background is created from data that has been smoothed by a \verb|SciPy| Gaussian kernel-density algorithm using Scott's rule \citep{scott_multivariate_1992} to determine the kernel size, and mirrored along the $V_Z=0$ axis because the background velocity distribution is here assumed to be approximately symmetric. 
\end{enumerate}
In panels b and c we can see that this background still contains some structure from the data and that the spiral pattern in panel c is not clear. 
\begin{enumerate}[resume]
    \item Find the spiral perturbation (Eq.~\ref{eqn:spiral}) that, multiplied by this background fits the data best (panel d). The parameter space is explored and the best fit is found by using a Markov Chain Monte Carlo (MCMC) \footnote{The model is implemented in Python using the package \texttt{emcee} \citep{foreman-mackey_emcee_2013} as an MCMC sampler.}
    approach. To find a fit, we need to define a probability function of a given model that takes the data and our prior into account. Given that we are using an MCMC sampler we can ignore any multiplicative constants and can say that the relevant value is $p$, where  
\begin{equation}
    \ln p = -\frac{1}{2} \sum \left( \frac{(\mathrm{mask}(N - f(r, \theta) \cdot B))^2}{f(r, \theta) \cdot B} \right) + \ln(P_{\mathrm{prior}}),
\label{eqn:log_func}
\end{equation}
    where $N$ is the data in the form of number counts for each bin, $B$ is the background, and $P_\mathrm{prior}$ is the prior probability. This perturbation is multiplied by the background and the mask (Eq.~\ref{eqn:outer_mask}) and compared to the data (panel e).
    \item Divide the data by the spiral perturbation produced in the fit to create an improved background which lacks some of the structure of the initial one (panel f). This new background is smoothed by averaging every pixel with its nearest neighbours (including diagonals) and is not necessarily symmetric in $V_Z$ anymore. The lack of imposed symmetries makes the smoothing very important as it prevents a solution where the background fits the data directly, bypassing the perturbation and driving the value of $\alpha$ to zero. Removing the requirement of symmetry also allows the entire distribution to drift in $V_Z$. This is not a concern as \cite{dehnen_distribution_1999} showed that the $V_Z$ distribution of stars in the Galactic disc is not symmetric about $V_Z = 0$. In fact, we see this mismatch between the ``symmetric-around-zero''-assumption and the data in panel c. 
\end{enumerate}
The process from point 3 to here is repeated until the new background no longer provides a better fit. The background converges quickly, usually not improving further after three iterations. The difference this process makes for the background can be seen by comparing panels c and g and noting in panel g the clearer spiral pattern. 
\begin{enumerate}[resume]
    \item When making a new background no longer improves the fit, take the final background and perturbation and make the final best fit (panel g). The final parameters are the median of the final samples found by the MCMC sampler. 
\end{enumerate}

The quality of the fit can be evaluated by looking at the residuals, computed by subtracting the best-fit phase spiral (panel h) from the data (panel a). Panel i shows these residuals. We can see that a weakness in the model is that the fitted spiral has fewer stars in the troth compared to the data as shown by the positive (red) spiral in panel i. This panel also highlights the differences between the mathematically perfect fitted spiral and the natural spiral from the data. 

The model is robust and capable of fitting spirals even to regions with relatively few stars. This is because the quality of the fit is judged by how well the smooth background is made to look like the data, and the spiral perturbation is the way in which it can change this background. 
In Fig.~\ref{fig:bad_example} we show an example of the process when dust severely obscures the sample. The figure contains data in the $8 < R / \mathrm{kpc} < 12$, $ 150^\circ < \phi < 155^\circ$, and $2200 < L_Z/$\,kpc\,km\,s$^{-1} < 2400$ ranges. The model still produces a reasonable fit and provides the parameters of the phase spiral. The residuals shown in the bottom panel display no spiral pattern and have generally low values, further indicating a good fit. 

\subsection{Rotation of the phase spiral} \label{ssection:rotation}

\begin{figure}
\centering
\includegraphics[width=0.85\hsize]{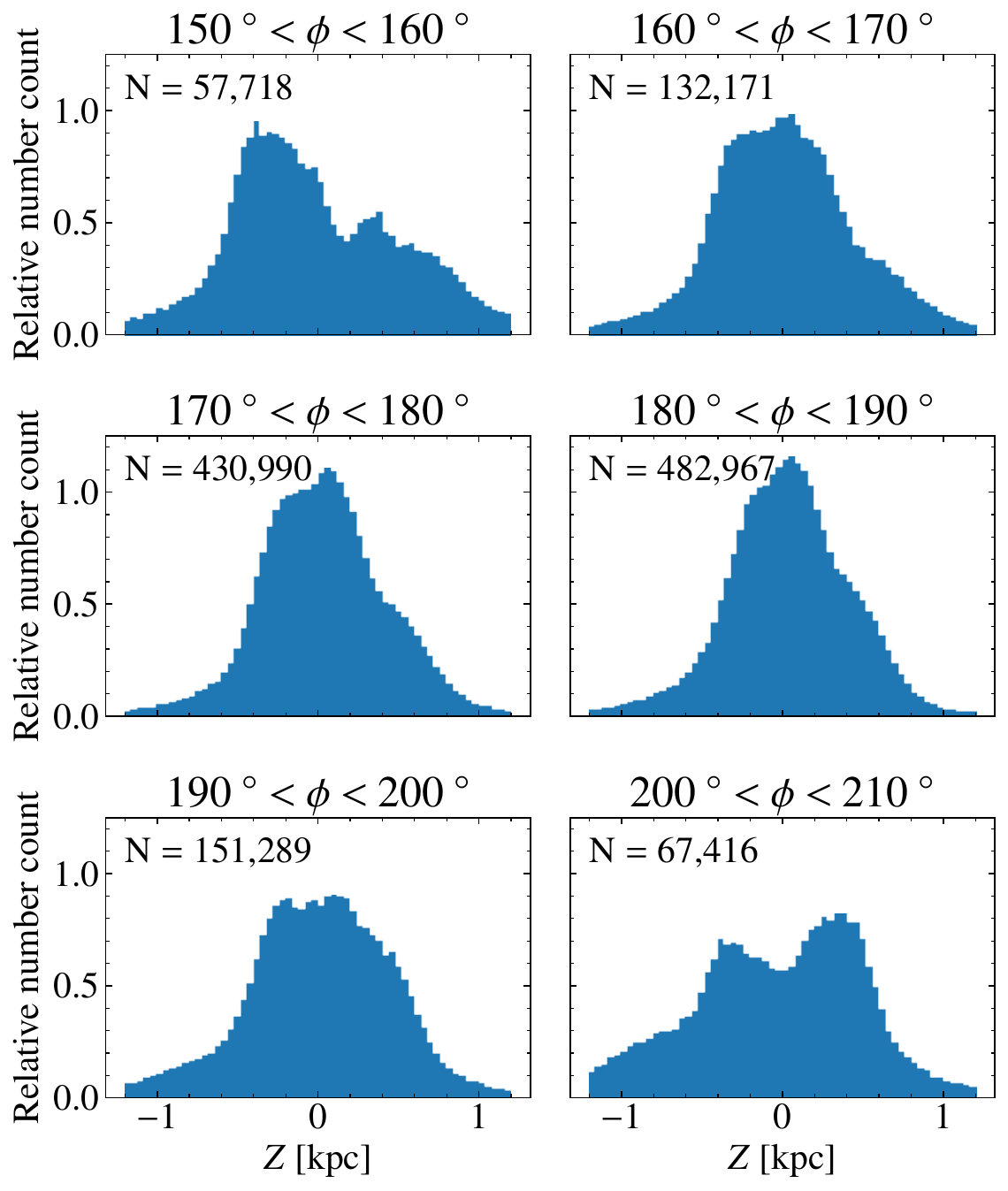}
  \caption{Normalized $Z$ distributions for stars at $2200 < L_Z/$\,kpc\,km\,s$^{-1} < 2400$ at different Galactic azimuths. Note the seemingly bimodal distribution at Galactic azimuth far from $180^\circ$, this is an effect of dust hiding stars in the middle of the Galactic disc. 
        }
     \label{fig:z_hist_grid}
\end{figure}

\begin{figure*}
\centering
\includegraphics[width=\hsize]{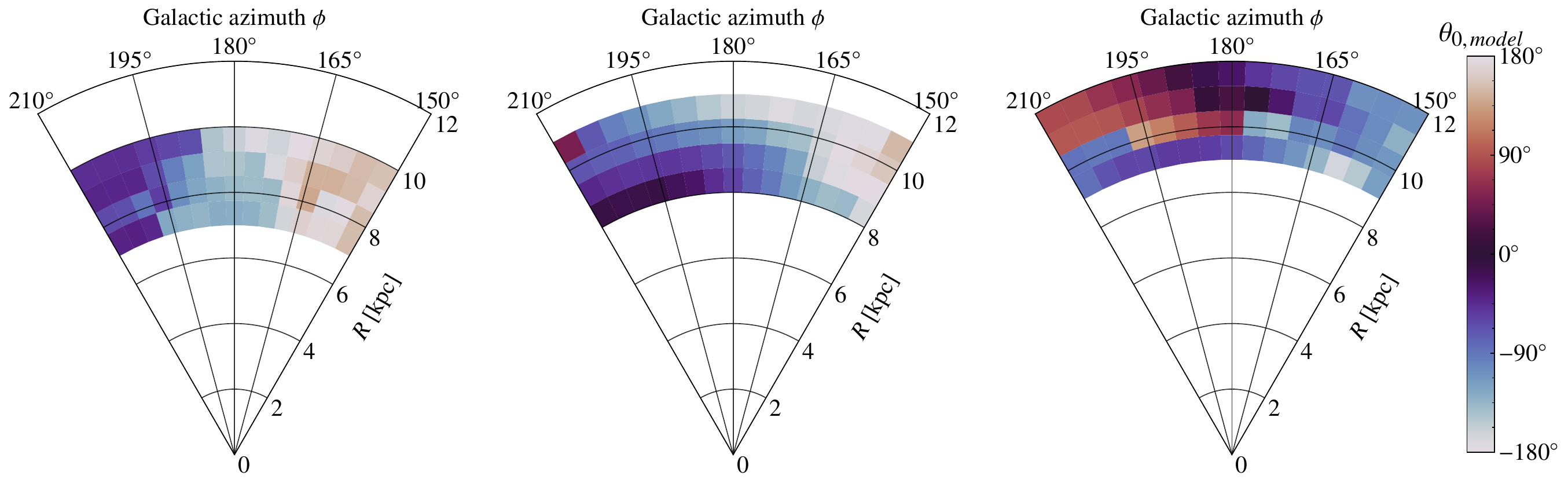}
  \caption{Angle ($\theta_{\,0,\:\mathrm{model}}$) of the phase spiral as measured by the model, showing the rotation across the Galactic disc. The plots show data for different regions of the Galaxy as seen from above for the three angular momentum ranges marked in Fig.~\ref{fig:L-plot}. The colour bar is periodic and the zero point is arbitrary. 
        }
     \label{fig:theta_tripple_wedge}
\end{figure*}

The animation\footnote{View the animation at \\\url{https://zenodo.org/records/12578839/files/medium_LZ_phase_spiral.gif?download=1}
} shows the phase spiral smoothly transition from stars at $\phi \approx 210^\circ $ to stars at $\phi \approx 150^\circ $ with bins $10^\circ$ wide. Here the phase spiral can be seen to spin clockwise about half a rotation as we decrease the Galactocentric azimuthal angle from $\phi = 150^\circ$ to $\phi = 210^\circ$. At either end of the range, there is a reduction of the number counts of stars in the mid-plane of the Galactic disc, because interstellar dust blocks our view of these stars. 
Figure \ref{fig:spiral_spin} shows the phase spirals at three different Galactic azimuths for stars with $2200 < L_Z/$\,kpc\,km\,s$^{-1} < 2400$, along with the perturbations fitted to the data. 
It is evident that the rotation angle of the phase spiral increases (rotating counterclockwise) with Galactic azimuth, changing by roughly $80^\circ$ over the $30^\circ$ change in azimuth from ${\sim}165^\circ$ to ${\sim}195^\circ$. 

The parameter $\theta_0$ which we fit in our model is not a convenient or particularly helpful description of the rotation of the phase spiral because the angle parameter ($\theta_0$) in the model has a degeneracy with the winding parameter ($b$) and to a certain extent the quadratic winding parameter ($c$) and different sets of these values can produce very similar spirals except in the most central regions, which are removed by the flattening function. 
Therefore, we describe the rotation of the phase spiral by the angle which maximises Eq.~\ref{eqn:spiral} (i.e. $\Delta\theta = 0$) at a fixed phase distance of $r = 500$\,pc and call this angle $\theta_{\,0,\:\mathrm{model}}$. This angle is shown in Fig.~\ref{fig:spiral_spin} with a red line, and the phase distance is shown with a white ring (scaled to the same axis ratio as the phase spiral) in the lower row. The angle $0^\circ$ is shown with a dashed white line in the upper row. 
Changing the phase distance at which we do these measurements does not change our results significantly except by changing all angles by some amount. For example, picking $r = 150$\,pc instead would shift all results by ${\sim}120^\circ$.

Figure~\ref{fig:z_hist_grid} shows the $Z$-distribution for six different ranges of Galactic azimuth, each $10^\circ$ wide, between $150^\circ$ and $210^\circ$, for stars in the $2200 < L_Z/$\,kpc\,km\,s$^{-1} < 2400$ range. Here, we can clearly see the reduction in the number of stars close to the Galactic plane ($Z \approx 0$) at high and low Galactic azimuth. This is because of dust in the plane of the Galactic disc, obscuring the true distribution of stars. Despite this, we see a shift as Galactic azimuth increases with more stars with low $Z$ at low Galactic azimuth than at higher Galactic azimuth, where there are more stars at high $Z$ instead. This is because stars in the phase spiral get pushed to greater $Z$ distances.

Figure~\ref{fig:theta_tripple_wedge} shows a map of the phase spiral's rotation angle ($\theta_{\,0,\:\mathrm{model}}$) on a top-down radial grid of the Galactic disc. Three plots are shown, each containing stars in a different angular momentum and Galactocentric radial distance range. The left plot contains stars in the $2000 < L_Z/$\,kpc\,km\,s$^{-1} < 2200$ and $7.5 < R / \mathrm{kpc} < 10$ range, the middle plot contains stars in the $2200 < L_Z/$\,kpc\,km\,s$^{-1} < 2400$ and $8.5 < R / \mathrm{kpc} < 11$ range and the right plot contains stars in the $2400 < L_Z/$\,kpc\,km\,s$^{-1} < 2600$ and $9.5 < R / \mathrm{kpc} < 12$ range, all between $150^\circ$ and $210^\circ$ in Galactic azimuth. The zero-point of the rotation angle is set to be $0$ at the $V_Z = 0$ line at $Z > 0$ (the positive $x$-axis as indicated in Fig.~\ref{fig:spiral_spin}). 
In Fig.~\ref{fig:theta_tripple_wedge}, we see a change in this rotation angle from high to low Galactic azimuth. In the left and middle plots, we see a relatively smooth decrease in rotation angle as Galactic azimuth decreases, changing by about $180^\circ$ over $60^\circ$ in the Galactic azimuth. The right panel shows the same trend but less smoothly. The left panel shows a radial increase in rotation angle by about $40^\circ$ over 2.5\,kpc while the middle panel shows a radial decrease in this angle by about $70^\circ$ over 2.5\,kpc. The right panel appears to show an increase in angle with radial distance. 

\subsection{Amplitude of the phase spiral} \label{ssection:strength}
\begin{figure*}
\centering
{\includegraphics[width=\hsize]{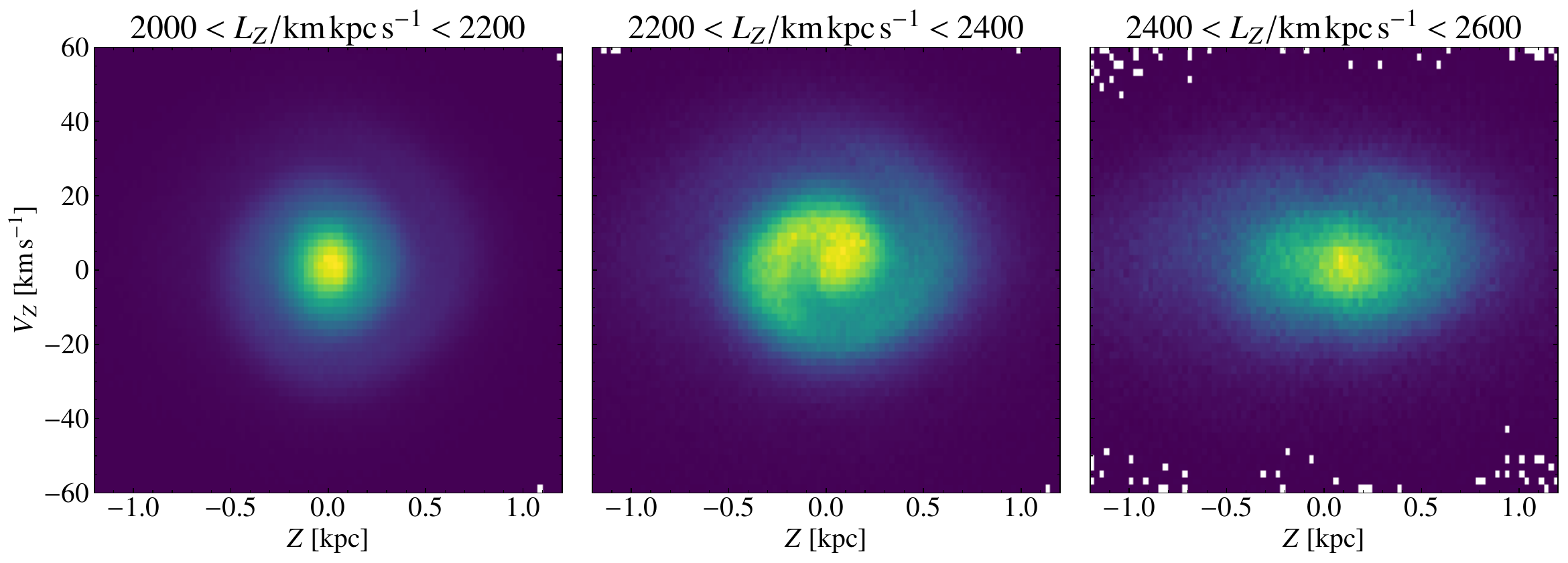}}\\
{\includegraphics[width=1.01\hsize]{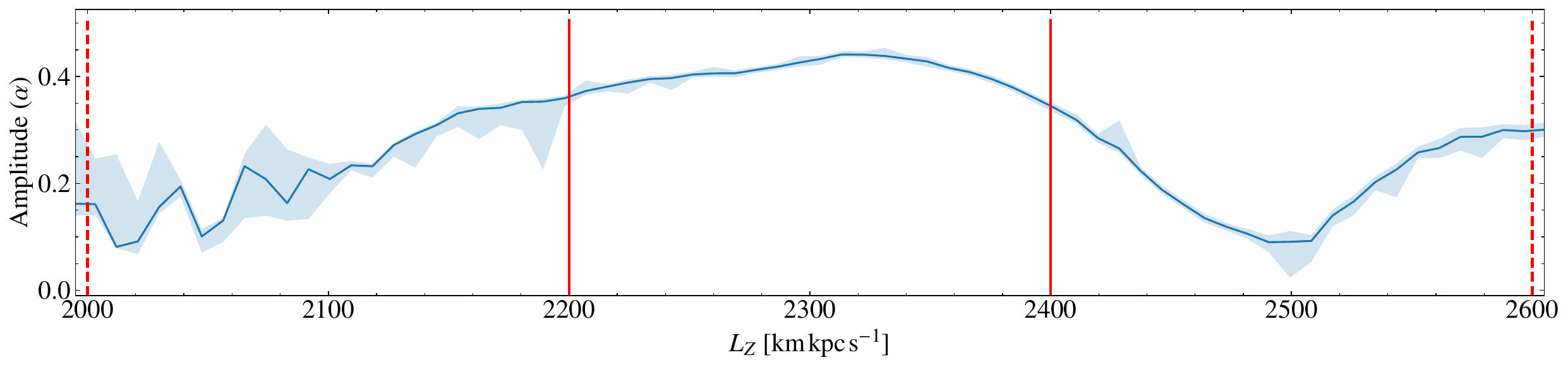}}
\caption{Measurements of the amplitude of the phase spiral as a function of angular momentum. Top: Number density of stars at low, medium, and high angular momentum, showing the phase spiral change shape and amplitude. Bottom: Amplitude of the phase spiral pattern as a function of angular momentum. The lines are the same as in Fig.~\ref{fig:L-plot}. The sample only includes stars that are within 500\,pc radially of where a star with the same angular momentum on a circular orbit would be and have a Galactocentric radial velocity of less than 22.5\,km\,s$^{-1}$ in order to restrict the selection to stars on cold orbits. The shaded area shows the 84th and 16th percentiles. 
        }
\label{fig:phase_spiral_and_alpha_LZ}
\end{figure*}

\begin{figure*}
\centering
\includegraphics[width=\hsize]{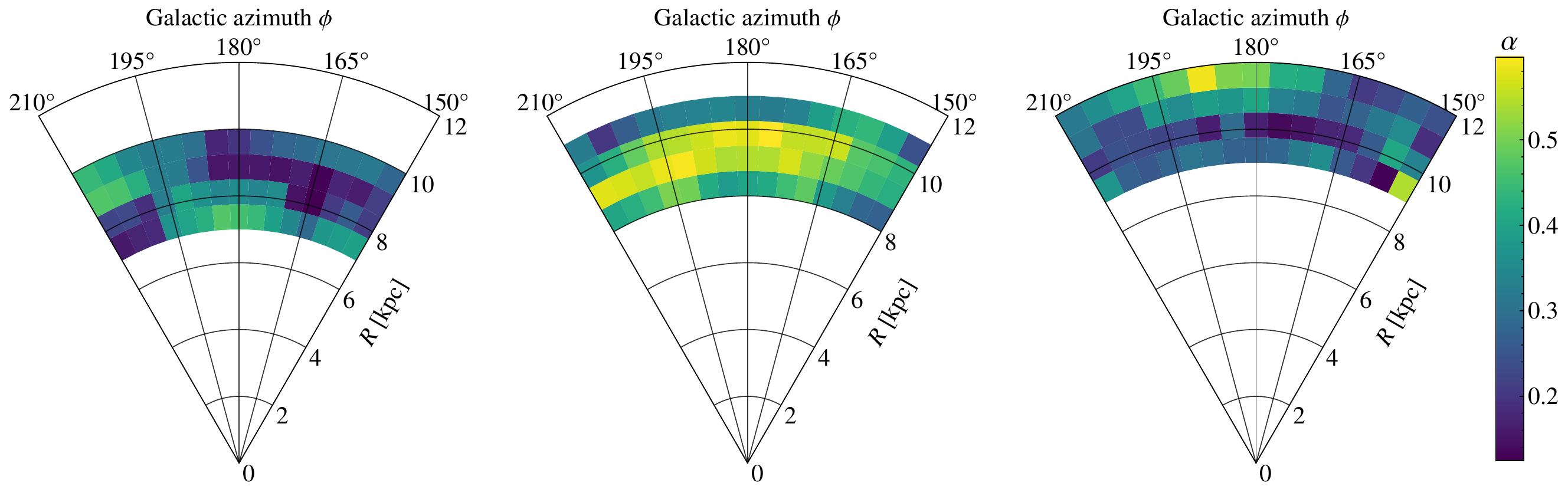}
    \caption{Amplitude ($\alpha$) of the phase spiral as measured by the model for different regions of the Galaxy as seen from above for three angular momentum ranges marked in Fig.~\ref{fig:L-plot}. The brightness of the plots corresponds to the height of the line in the bottom panel in Fig.~\ref{fig:phase_spiral_and_alpha_LZ}, showing the change in amplitude across the Galactic disc. 
        }
    \label{fig:alpha_tripple_wedge}
\end{figure*}

Figure~\ref{fig:phase_spiral_and_alpha_LZ} shows the $Z$-$V_Z$ phase plane for the three regions marked with lines in Fig.~\ref{fig:L-plot}. The left and right panel contains stars in the $2000 < L_Z/$\,kpc\,km\,s$^{-1} < 2200$ and $2400 < L_Z/$\,kpc\,km\,s$^{-1} < 2600$ ranges respectively. Both these regions show weak and/or almost dissolved phase spiral patterns. The middle panel, which corresponds to the $2200 < L_Z/$\,kpc\,km\,s$^{-1} < 2400$ range, shows a clear, single-armed, phase spiral pattern. 

Our model contains a parameter for the amplitude, or strength, of the phase spiral pattern ($\alpha$). 
Figure \ref{fig:phase_spiral_and_alpha_LZ} shows the amplitude of the phase spiral as a function of angular momentum in the bottom panel. There is a peak at $L_Z \approx 2300$ kpc\,km\,s$^{-1}$, which is what we expected from Fig.~\ref{fig:L-plot} and the top row of Fig.~\ref{fig:phase_spiral_and_alpha_LZ}. The shaded region in the plot is between the 84th and 16th percentiles. These are used to show the statistical uncertainties in the model. The systematic uncertainties are expected to be larger. The jagged part between $L_Z \approx 2000$\,kpc\,km\,s$^{-1}$ and $2100$\,kpc\,km\,s$^{-1}$ are examples of the modelling procedure finding an alternate solution. By visual inspection, we can conclude that the phase spirals found at these points are not the best fits. 
The line rises at the high end of the plot, indicating another peak at or beyond $L_Z = 2600$\,kpc\,km\,s$^{-1}$. This seems to correspond to ``$V_Z$ feature 2'' in Fig.~\ref{fig:L-plot} and the bifurcation discussed by \cite{mcmillan_disturbed_2022}.
The data in Fig.~\ref{fig:phase_spiral_and_alpha_LZ} is limited to stars with less than $ |22.5|$\,km\,s$^{-1}$ in Galactocentric radial velocity, which represents $\sigma/2 $ (one half standard deviation) of the velocity distribution. In this way, we can select only stars on dynamically cold orbits (close to circular). We do this because we want to compare our results against those of \cite{li_dissecting_2020} who investigate the strength of the phase spiral in stars on hot or cold orbits in the solar neighbourhood. The bottom plot contains points based on bins that are 1000\,pc by $30^\circ$. These bins are centred on the guiding centre radius corresponding to that angular momentum. Because the bins are $30^\circ$ wide, we are measuring phase spirals with rotational angles over a ${\sim}70^\circ$ range, as large as that seen in Fig.~\ref{fig:spiral_spin}. 

Figure~\ref{fig:alpha_tripple_wedge} shows a map of the amplitude ($\alpha$) on a top-down radial grid of the Galactic disc. Three plots are shown, each containing stars in a different angular momentum and Galactocentric radial distance range, the same as in Fig.~\ref{fig:theta_tripple_wedge}. The figure shows that the brightest region, with the highest amplitude, is the middle panel with stars in the $2200 < L_Z/$\,kpc\,km\,s$^{-1} < 2400$ range as we would expect from Fig.~\ref{fig:phase_spiral_and_alpha_LZ}. The figure also shows that the region of the highest amplitude moves outward with $L_Z$, as well as in each bin. There is a slight trend for the amplitude to decrease at higher and lower Galactic azimuths in the figure. This is believed to be an observational effect that comes from the much greater distances to those areas. 

\subsection{Chemical composition of the phase spiral}\label{ssection:chem}
\begin{figure*}
\centering
\includegraphics[width=\hsize]{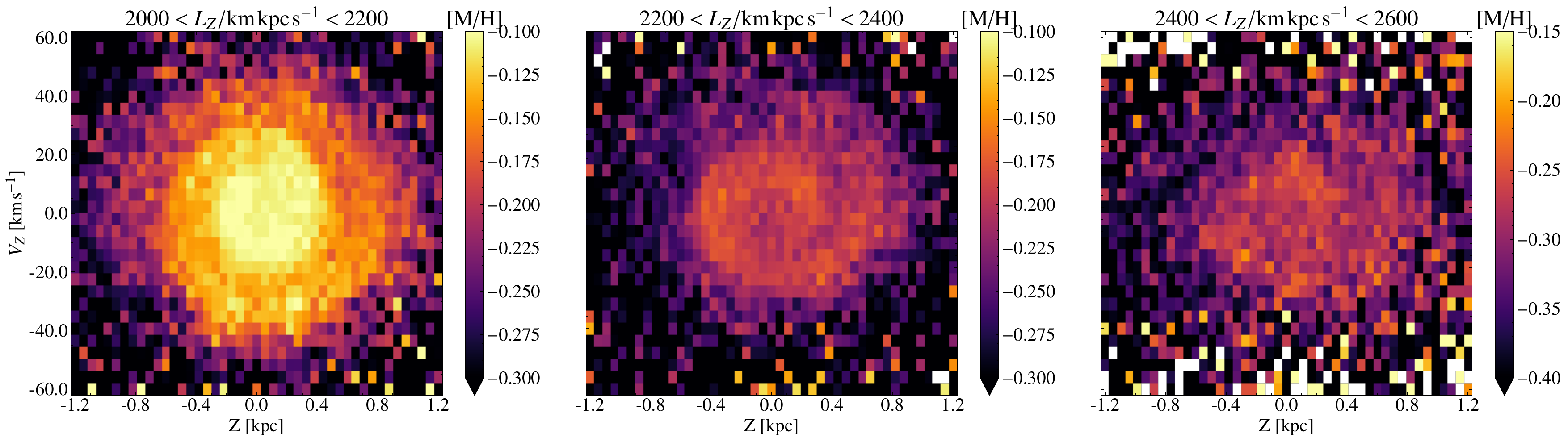}
  \caption{Phase spirals coloured by mean global metallicity at low, medium, and high angular momentum, showing that the spiral pattern is visible, compare to Fig.~\ref{fig:phase_spiral_and_alpha_LZ}. Note that the rightmost panel has different values on the colour bar. The data is split into the three angular momentum ranges marked in Fig.~\ref{fig:L-plot}.
  }
     \label{fig:metal_spiral}
\end{figure*}

Figure~\ref{fig:metal_spiral} shows the $Z$-$V_Z$ phase plane coloured by the mean global metallicity for the sample of stars that we have metallicities for in the same three ranges in angular momentum as in Fig.~\ref{fig:phase_spiral_and_alpha_LZ}. A similar, but weaker, spiral pattern can be seen here. We observe that stars in the phase spiral have a slightly higher metallicity than those outside the pattern, indicating a common origin between the stars in the arm of the phase spiral and those in the Galactic thin disc. 
A clear decreasing trend in mean metallicity with angular momentum can also be seen. 
The similarities between the spiral patterns in Figs.~\ref{fig:phase_spiral_and_alpha_LZ} and \ref{fig:metal_spiral} are noteworthy but not surprising as stars in the Galactic thin disc are known to have higher metallicity and the phase spiral is assumed to be a perturbation of the Galactic disc which moves stars away from the midplane. They both show stars in the same angular momentum ranges and the same phase spiral pattern appears. In the left panel, the central region of phase space (the thin disc) shows high [M/H] values. In this panel, an arm of the phase spiral can be seen emerging from the top of this central region, at about $Z \approx  -300$\,pc, $V_Z \approx 20$\,km s$^{-1}$. In the middle panel, a one-armed spiral is visible in stars with mean metallicity of $\langle \mathrm{[M/H]} \rangle \approx-0.15$ against a background of $\langle \mathrm{[M/H]} \rangle \approx-0.22$. This panel lacks the high metallicity region in the centre of the phase plane, instead having the region of highest metallicity be in the arm of the phase spiral. Even the less dense gap between the wraps of the phase spiral arm is visible as a darker curve near the centre of the phase plane. In contrast to the left panel, this one does not have the highest value in the very centre of the phase plane but shows a small decrease instead. The right panel shows a faint trace of a spiral arm at $Z \approx 500$\,pc, $V_Z\,{\approx}\,-20$\,km s$^{-1}$. Note that the colour scale in this panel is shifted slightly towards lower metallicity values in order to bring out the remaining structure.

\section{Discussion}\label{section:discussion}

\subsection{Formation}
The most established theories for the formation of the phase spiral indicate a single-impact formation mechanism, caused by the Sagittarius dwarf galaxy. However, numerous recent papers are pointing in the opposite direction, that a single-impact origin scenario is too simple to explain all the observations \citep[e.g.][]{tremaine_origin_2023, frankel_vertical_2023}. 
The phenomena presented in this paper challenge certain proposed formation mechanisms for the phase spiral. Any model of the phase spiral must be able to reproduce the smoothly changing angle of the phase spiral across a wide range of different Galactic azimuths and radii, such as that which we show in Fig.~\ref{fig:theta_tripple_wedge} and our animation\footnote{\url{https://zenodo.org/records/12578839/files/medium_LZ_phase_spiral.gif?download=1}
}. The observed change of the angle would intuitively fit with a single-impact formation scenario. 

\cite{garcia-conde_phase_2022} look at phase spirals in a cosmological simulation and conclude that phase spirals  appear even if the interacting satellite galaxies are less massive or more distant than the Sagittarius dwarf galaxy is thought to have been \citep{Niederste-Ostholt_re-assembling_2010}. 
This raises the interesting possibility of these simulations bringing clarity into whether a multi-impact formation scenario can produce a smoothly changing phase spiral. Current simulations do not provide the resolution to answer this question, possibly because they are not set up to. While \cite{garcia-conde_phase_2022} do show phase spirals for different Galactic azimuths, the scale is such that the entire area considered in this paper fits in two bins (see their Figs.~2 and 5 for details). 

Knowledge of how the phase spiral shifts across the Galactic disc can be related to the properties of the Galactic disc and the cause of the perturbation. \cite{widmark_weighing_2021} used the velocities of stars in the Galactic disc and phase spiral to infer the potential of the Galactic disc, and thereby its mass. Comparing how the phase spiral has propagated through the Galactic disc with results from modelling studies can lead to better constraints for these methods. It should be noted that, as found by many authors \citep[e.g.][]{gomez_fully_2016, laporte_influence_2018, grand_ever-present_2023}, the torque on the Galactic disc caused by the dark matter wake of a passing satellite galaxy can be significantly stronger than the direct interaction with the satellite galaxy. This means that the connection between the perturbing satellite galaxy and the perturbation in the Galactic disc may not be as simple as some previous idealised models have assumed. Explaining the physics on scales as small as those focused on in this paper (roughly the area covered by Fig.~\ref{fig:theta_tripple_wedge}) in the context of a cosmological simulation presents a challenge for the modelling community.

\subsection{Hot/cold orbits} 
\cite{li_dissecting_2020} and \cite{bland-hawthorn_galah_2019} argue that the phase spiral is more clearly defined in stars on dynamically cold (close to circular) orbits than in stars on dynamically hot (far from circular) orbits. \cite{li_dissecting_2020} specifically argue that stars on hotter orbits should be excluded from samples used in phase-mixing studies to provide clearer samples. 
The results of this paper combined with those of \cite{frankel_vertical_2023} are in tension with these conclusions. Figure~\ref{fig:phase_spiral_and_alpha_LZ} consists of a subset of the full selection of stars. It contains stars on cold orbits by only including stars that are within 500\,pc radially of where a star with the same angular momentum on a circular orbit would be and have a Galactocentric radial velocity of less than $|22.5|$\,km\,s$^{-1}$, which is approximately $ \sigma/2 $ of the Galactocentric radial velocity. The strength of the phase spiral in these kinematically cold stars is similar to that by \cite{frankel_vertical_2023}, who show that stars on hot orbits at $L_Z \approx 2300$\,kpc km s$^{-1}$ still produce a phase spiral with higher amplitude than stars on cold orbits at $L_Z \approx 1800$\,kpc km s$^{-1}$ (see their Fig.~8 for details). Both results show the same feature, a region with a more prominent phase spiral, despite containing separate populations of stars with different dynamics. The conclusion we draw is that the phase spiral and its connection to the Galaxy is more complicated than can be evaluated based on one parameter.

The investigation of the amplitude of the phase spiral pattern as a function of angular momentum in the range $1250 < L_Z/$\,kpc\,km\,s$^{-1} < 2300$ conducted by \cite{frankel_vertical_2023} is similar to ours but there are some key differences. 
Their sample consists of stars within a 0.5\,kpc cylinder centred on the Sun meaning that the stars included in this volume with high angular momentum ($>2000$\,kpc\,km\,s$^{-1}$) are all going to be on relatively dynamically hot orbits. Our sample contains stars whose position and guiding centre are further out, meaning that when considering the high angular momentum case, the stars are on dynamically cooler orbits. 
They show a general increase in amplitude with angular momentum, with the highest peak at $L_Z \approx 2350 $\,kpc\,km\,s$^{-1}$, (see their Fig.~8 for details). 
The bins containing stars with high angular momentum in their sample hold few stars leading to a relatively large scatter in the results. Our results also extend to higher angular momentum meaning that, in Fig.~\ref{fig:phase_spiral_and_alpha_LZ}, we can see the dip in amplitude at $L_Z \approx 2500 $ kpc\,km\,s$^{-1}$.

The questions posed in \cite{bland-hawthorn_galah_2019} are still relevant. How are different populations of stars affected by whatever mechanism caused the phase spiral? How is the gas affected? Were the stars in the phase spiral formed in it or were they swept up into it after they had already formed? These questions are mostly outside the scope of this paper but could bring significant insights into the dynamic processes that shape our galaxy. 

\subsection{Metallicity}
\cite{widrow_galactoseismology_2012} discovered an asymmetry in the $Z$-distribution of stars in the Galactic disc which we now associate with the phase spiral. They found that when looking at the number density of stars as (North $-$ South) / (North $+$ South) the result is $<0$ at $|Z| \,{\approx}\, 400$\,pc and $>0$ at $|Z| \,{\approx}\, 800$\,pc. \cite{an_asymmetric_2019} analysed this asymmetry further, specifically looking at the metallicity of the stars. They found that the vertical metallicity distribution is asymmetric in a complicated manner similar to the number density.
Our results suggest that the arm of the phase spiral drives stars to greater $ Z$-distances in the region of the Galaxy we study. This would push stars from the Galactic disc vertically away from it, and preferentially in the positive $Z$-direction \citep{an_asymmetric_2019}. We can see this in Fig.~\ref{fig:metal_spiral} where the phase spiral is shown in metallicity and causes the positive $Z$-side of each plot to be more metal-rich at large distances than the negative $Z$-side. 
 
\cite{bland-hawthorn_galah_2019} looked at the difference in the phase spiral when using different cuts in the elemental abundance plane. They found that more metal-rich ($\rm [Fe/H]>0.1$) stars were concentrated in the central part of the phase spiral. As we can see in Fig.~\ref{fig:metal_spiral}, we also see that stars with higher mean metallicity can be found in the centre of the phase spiral for stars with low angular momentum (left panel). The rest of the panels indicate that the most metal-rich stars exist not close to $0$\,kpc, $0$\,km\,s$^{-1}$ but in the arm of the phase spiral at a distance of $\approx 0.5$\,kpc. The conclusion is that these stars were formed in the Galactic thin disc and then perturbed to move out of it. This would explain the asymmetry in the $ Z$-distribution and the concentration of metal-rich stars in the phase spiral. 

\subsection{Effects of rotation}
If the rotation of the phase spiral is not taken into consideration when studying it, some features are at risk of being washed out. For example, in Fig.~\ref{fig:L-plot}, the sample is restricted to stars with Galactic azimuth of $175^\circ < \phi < 185^\circ$, otherwise, the feature of interest is not clearly visible. Future authors should be aware of this phenomenon and how it may affect their results. 

In Fig.~\ref{fig:z_hist_grid} it appears that stars are missing in the centre of the Galactic disc at high or low Galactic azimuth. This is attributed to dust. We also see an asymmetry in the $ Z$ distribution when comparing regions at high and low Galactic azimuth. This effect could be caused by the rotation of the phase spiral as it brings the phase spiral arm out of the high $Z$ region at lower Galactic azimuth. We do not believe this is caused by the warp of the Galactic disc, as the warp only starts being measurable at Galactocentric distances greater than those considered here, at about 10\,kpc \citep{cheng_exploring_2020}. 
However, it seems like the phase spiral and the Galactic warp overlap in certain regions of the Galaxy and are perhaps related. The peak in the phase spiral amplitude is found at $L_Z = 2300 $\,km\,kpc\,s$^{-1} $ which corresponds to a guiding radius of $R_g \approx 9.9$ kpc and reaches a minimum at $L_Z = 2500 $\,km\,kpc\,s$^{-1} $ which corresponds to $R_g \approx 10.8$ kpc which is inside of the distance where the Galactic warp starts being relevant \citep[e.g.][]{cheng_exploring_2020}. 

\section{Summary and Conclusions}\label{section:conclusions}
In this work, we use data from \textit{Gaia} DR3 to investigate the \textit{Gaia} phase spiral by making a new model capable of fitting several of its key characteristics. We use a sample of stars with measured radial velocities to get full three-dimensional information on both their position and velocity, a sample of about 31.5 million stars. Using our model, we have been able to determine the rate of rotation of the phase spiral with Galactic azimuth and the amplitude of the phase spiral as a function of angular momentum. We find that, for the data we explore, the phase spiral rotates with Galactic azimuth. We find a peak in the amplitude of the phase spiral at $L_Z {\approx} 2300$\,km\,kpc\,s$^{-1}$ which manifests as a very clear phase spiral pattern in number density when using only stars with similar angular momentum. 

Our main findings in this paper are listed here:
\begin{enumerate}
    \item The phase spiral changes orientation along both Galactic radial distance and Galactic azimuth, and it rotates at a rate which is three times the rate of the azimuthal angle, a rate of ${\sim} 180^\circ $ per $60^\circ$ Galactic azimuth for stars with angular momenta from 2000\,km\,kpc\,s$^{-1}$ to 2400\,km\,kpc\,s$^{-1}$, corresponding to orbits typically found outside the Sun's position in the Galaxy. 
    \item The amplitude of the phase spiral pattern changes with angular momentum with a peak at about $2300 \pm 100$ kpc\,km\,s$^{-1}$, producing a substantially clearer spiral pattern in number density.
    \item The stars in the phase spiral arm are chemically very similar to those in the $Z$-centre of the Galactic disc. This indicates that the stars in the phase spiral originally belonged to the Galactic thin disc. 
    \item We can confirm the conclusions of \cite{an_asymmetric_2019} and \cite{bland-hawthorn_galah_2019} that the Z-asymmetry of the metallicity gradient of the Galaxy is caused by the metal-rich arm of the phase spiral pushing such stars to greater $Z$-positions.
\end{enumerate}

The reason for the change in the $L_Z$-$V_Z$ distribution between the solid lines in Fig.~\ref{fig:L-plot}, the overdensity seen below the thick line, was found to be the phase spiral. In Fig.~\ref{fig:phase_spiral_and_alpha_LZ}, we show the number density of the phase spiral in three regions in the top row. Here we can see that the central panel, containing stars in the 2200\,km\,kpc\,s$^{-1}$ to 2400\,km\,kpc\,s$^{-1}$ range, shows a clearer and more defined spiral pattern. The bottom of the same figure shows the amplitude of the phase spiral as measured by the model. Here we also see that the phase spiral is strongest in the 2200\,km\,kpc\,s$^{-1}$ to 2400\,km\,kpc\,s$^{-1}$ range.
The central line in Fig.~\ref{fig:L-plot} is raised to about 15 km\,s$^{-1}$, corresponding to when the phase spiral first turns onto the negative $Z$-values, the lower clump sits at $-20$\,km\,s$^{-1}$ which corresponds to then the spiral arm turns back to the positive $Z$-values.

By combining the data from \textit{Gaia} with that coming from the soon-to-be operational spectrographs 4MOST \citep{bensby_4most_2019, chiappini_4most_2019} and WEAVE \citep{jin_wide-field_2023}, more light will be shed on the origins of the phase spiral by revealing detailed chemical abundances for millions of stars in all parts of the Milky Way.

\begin{acknowledgements}
    \newline PM gratefully acknowledges support from project grants from the Swedish Research Council (Vetenskapr\aa det, Reg: 2017-03721; 2021-04153). TB and SA acknowledge support from project grant No.~2018-04857 from the Swedish Research Council. Some of the computations in this project were completed on computing equipment bought with a grant from The Royal Physiographic Society in Lund.   
    This work has made use of data from the European Space Agency (ESA) mission
    {\it Gaia} (\url{https://www.cosmos.esa.int/gaia}), processed by the {\it Gaia}
    Data Processing and Analysis Consortium (DPAC,
    \url{https://www.cosmos.esa.int/web/gaia/dpac/consortium}). Funding for the DPAC
    has been provided by national institutions, in particular the institutions
    participating in the {\it Gaia} Multilateral Agreement. 
    This research has made use of NASA’s Astrophysics Data System. 
    This paper made use of the following software packages for Python,
    \verb|Numpy| \cite{harris_array_2020}, 
    \verb|AstroPy| \cite{astropy_collaboration_astropy_2022}, 
    \verb|emcee| \cite{foreman-mackey_emcee_2013}, 
    \verb|SciPy| \cite{virtanen_scipy_2020}.
\end{acknowledgements}

%
%

\bibliographystyle{aa}
\bibliography{paper1_update_ads2nty.bib}
\onecolumn

\begin{appendix}
\section{Further data}
\begin{figure*}[h]
Figure~\ref{fig:phi_grid_low} shows the phase spiral for 12 different ranges of Galactic azimuth, each $5^\circ$ wide, between $150^\circ$ and $210^\circ$, for stars in the $2000 < L_Z/$\,kpc\,km\,s$^{-1} < 2200$ range. Each panel has the rotation angle ($\theta_{\,0,\:\mathrm{model}}$) marked with a red line. Figs.~\ref{fig:phi_grid} and \ref{fig:phi_grid_high} show the same for the $2200 < L_Z/$\,kpc\,km\,s$^{-1} < 2400$ and $2400 < L_Z/$\,kpc\,km\,s$^{-1} < 2600$ ranges respectively. 
The angle of the phase spiral changes with Galactic azimuth in this figure as well, rotating about $200^\circ$ over an azimuth range of about $55^\circ$. The phase spiral is most clearly seen around $\phi = 180^\circ$ and gets less prominent in both azimuthal directions as the number of stars in our data decreases. 

\includegraphics[width=\hsize]{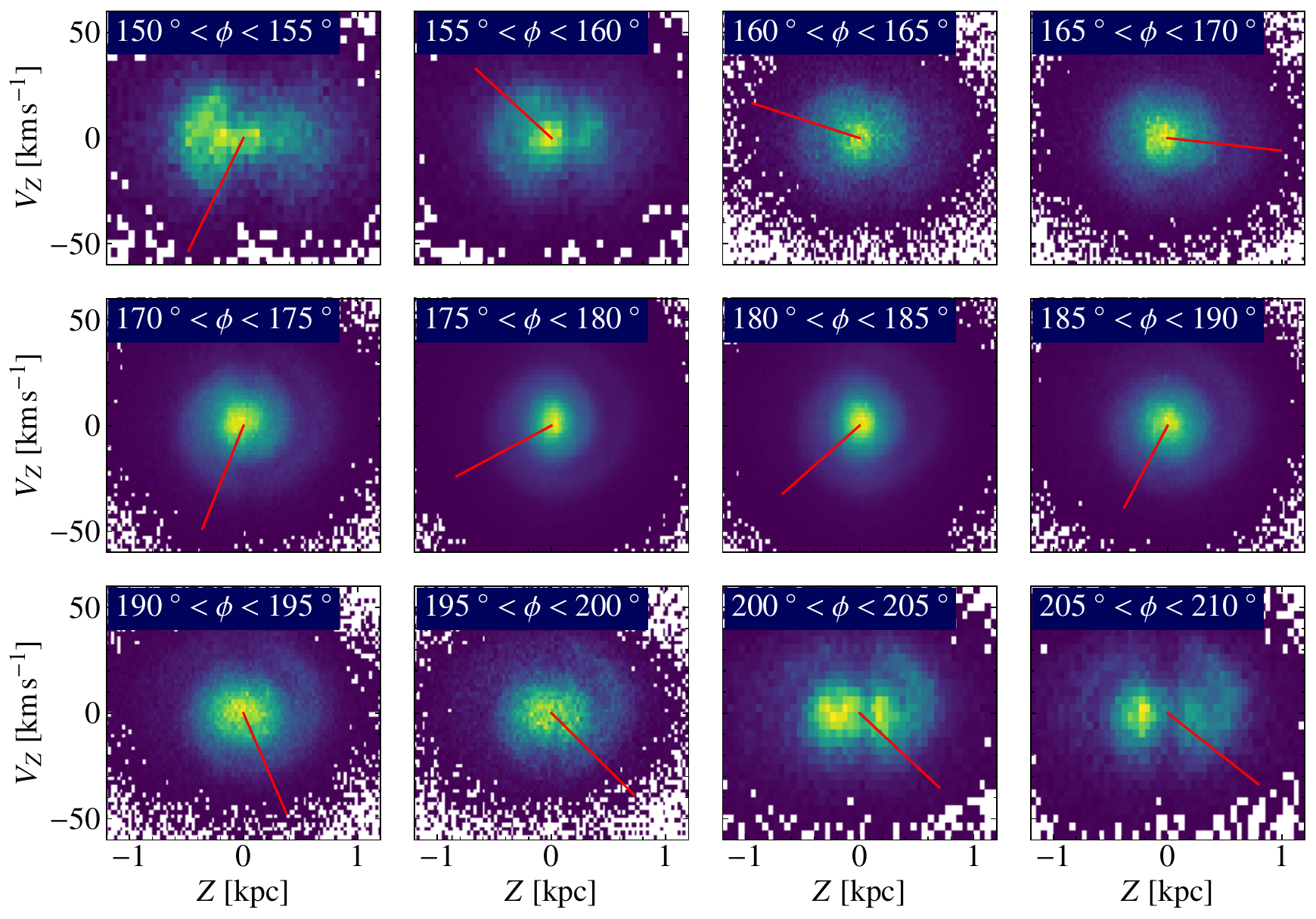}
  \caption{Phase spirals and fitted spiral perturbations at different values for Galactic azimuth ($\phi$) in the $2000 < L_Z/$\,kpc\,km\,s$^{-1} < 2200$ range. The measured angle of the phase spiral is marked with a red line. The azimuthal range of the stars is marked with text in the right panels. 
  This figure shows how we, by using the model, can determine the angle of the phase spiral even in regions where the data is of lower quality due to extinction by dust. 
        }
     \label{fig:phi_grid_low}
\end{figure*}

\begin{figure*}[ht]
\includegraphics[width=\hsize]{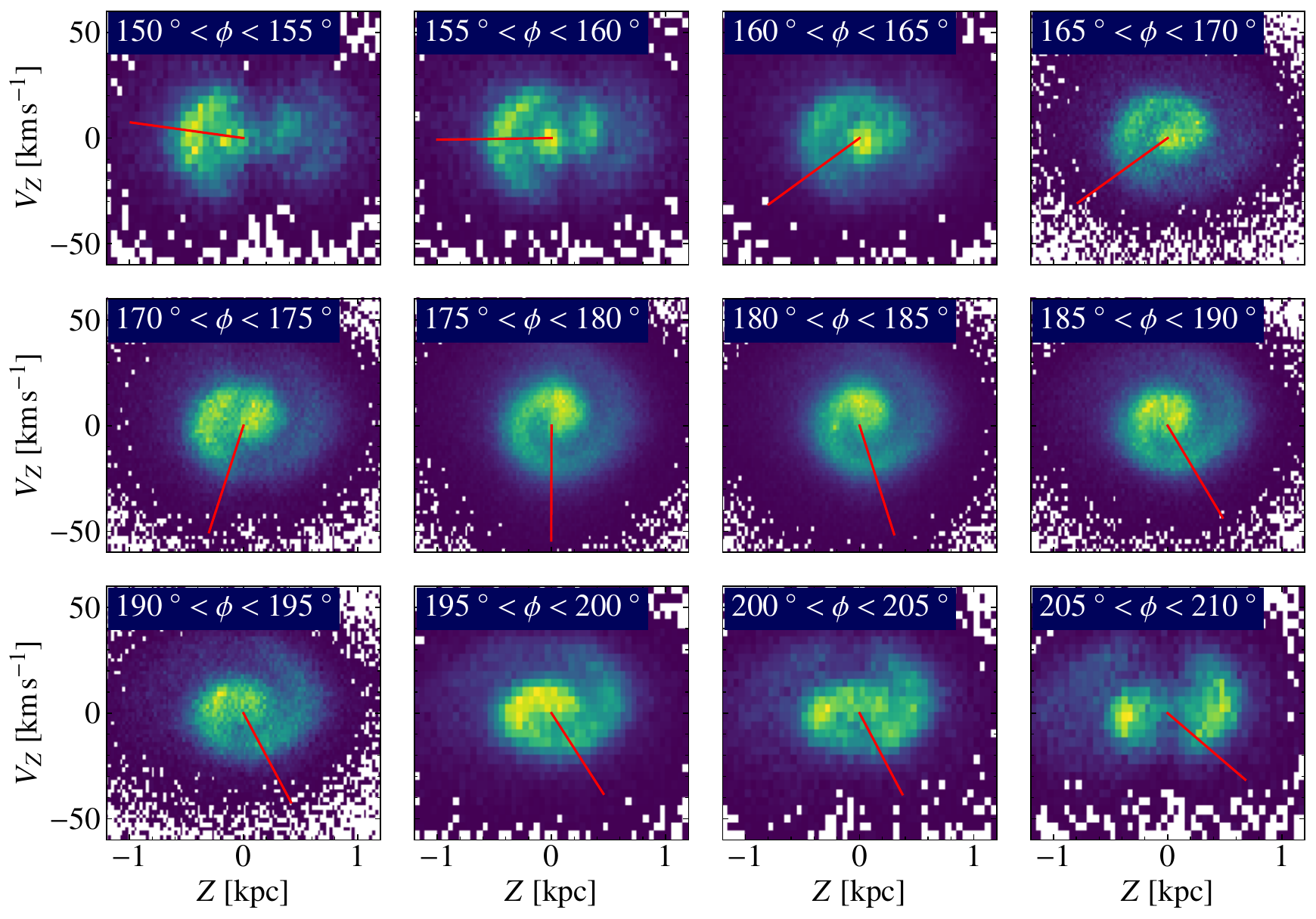}
  \caption{Phase spirals and fitted spiral perturbations at different values for Galactic azimuth ($\phi$) in the $2200 < L_Z/$\,kpc\,km\,s$^{-1} < 2400$ range. The measured angle of the phase spiral is marked with a red line. The azimuthal range of the stars is marked with text in the right panels. 
        }
     \label{fig:phi_grid}
\end{figure*}

\begin{figure*}[h]
\includegraphics[width=\hsize]{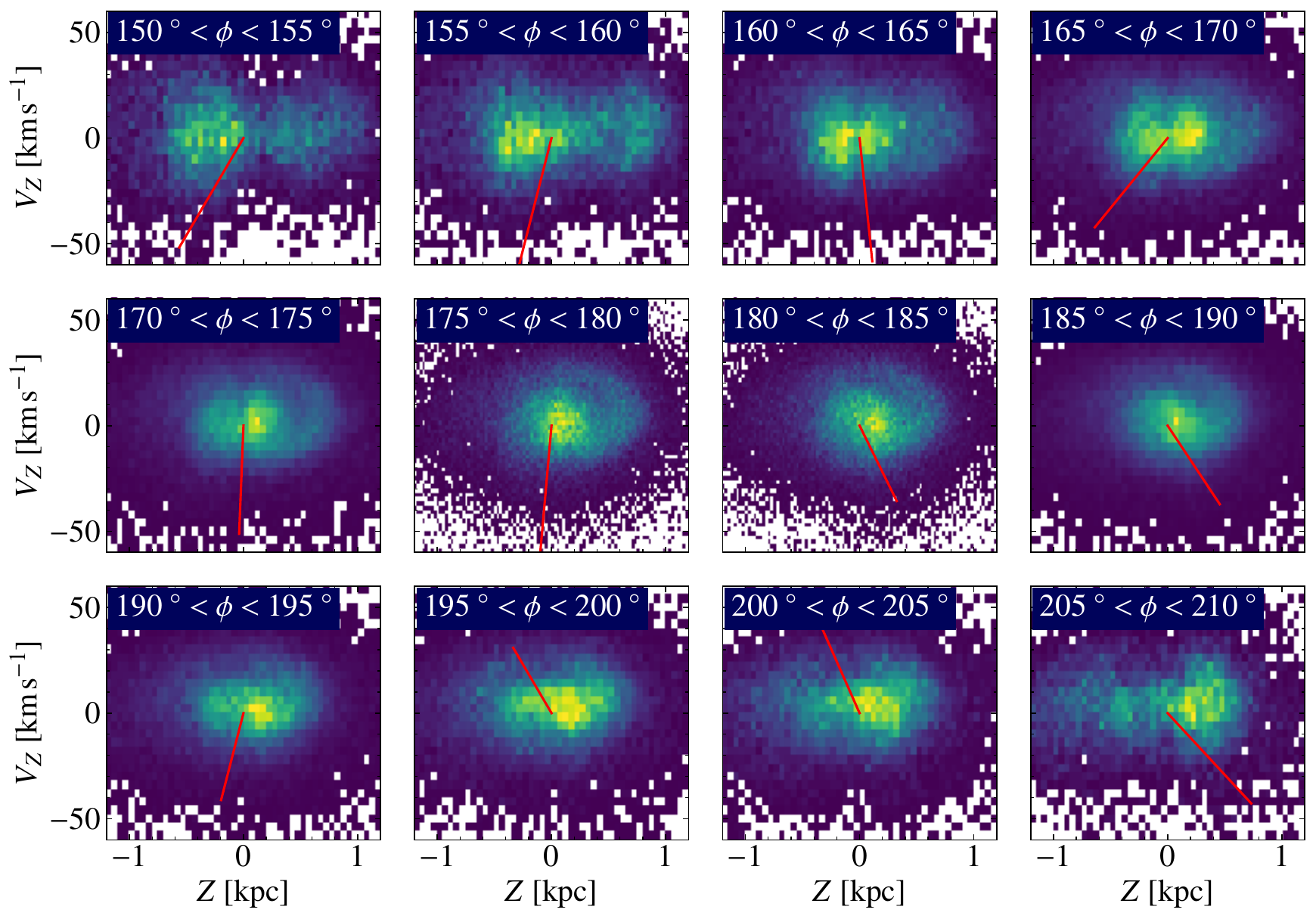}
  \caption{Phase spirals and fitted spiral perturbations at different values for Galactic azimuth ($\phi$) in the $2400 < L_Z/$\,kpc\,km\,s$^{-1} < 2600$ range. The measured angle of the phase spiral is marked with a red line. The azimuthal range of the stars is marked with text in the right panels. 
        }
     \label{fig:phi_grid_high}
\end{figure*}

Animations of the phase spiral over a range of Galactic azimuths, similar to the one earlier in the paper, for stars at lower and higher angular momentum, are available at \url{https://zenodo.org/records/12578839/files/low_LZ_phase_spiral.gif?download=1} and at \url{https://zenodo.org/records/12578839/files/high_LZ_phase_spiral.gif?download=1}. 

\end{appendix}

\end{document}